\documentclass[aps,prb,twocolumn,showpacs,10pt,floatfix,superscriptaddress,floatfix, longbibliography]{revtex4-1}
\usepackage{amsmath}
\usepackage{braket}
\usepackage{amssymb}
\usepackage[normalem]{ulem}
\usepackage{ dsfont }
\usepackage{bm}
\usepackage[dvipsnames]{xcolor}
\usepackage{stmaryrd}
\usepackage{bbm}
\usepackage{mathtools}

\newcommand{\vk}{{\vec k}}
\newcommand{\ZZ}{\mathbb{Z}}

\newcommand{\va}{\vec a}

\newcommand{\vn}{\vec n}

\newcommand{\vR}{\vec R}

\newcommand{\vx}{\vec x}

\newcommand{\vA}{\vec A}

\begin{document}

\title{$N$-band Hopf insulator}
\author{Bastien Lapierre, Titus Neupert, and Luka Trifunovic}
\affiliation{Department of Physics, University of Zurich, Winterthurerstrasse 190, 8057 Zurich, Switzerland}
\date{\today}

\begin{abstract}

We study the generalization of the three-dimensional two-band Hopf insulator to the case of
many bands, where all the bands are separated from each other by band gaps. The
obtained $\ZZ$ classification of such a $N$-band Hopf insulator is related to the quantized isotropic magnetoelectric coefficient of its bulk. The boundary of a $N$-band Hopf insulator can be fully gapped, and we find that there is no unique way of dividing a finite system into bulk and boundary. Despite this non-uniqueness, we find that the magnetoelectric coefficient of the bulk and the anomalous Hall conductivity of the boundary are quantized to the same integer value. We propose an experiment where the quantized boundary effect can be measured in a non-equilibrium state.
\end{abstract}

\maketitle   
\section{Introduction}
Topological materials exhibit robust boundary effects that promise many
applications. For example, more energy-efficient microelectronics can be designed by making use of backscattering-free edge modes, i.e., chiral (helical) modes appearing in the
quantum (spin) Hall systems, and similarly, the surface states of three-dimensional $\mathbb{Z}_2$ topological insulators can serve as a good catalyst.~\cite{chen2011} Another promising application is a fault-tolerant quantum computing
 based on Majorana zero-energy states appearing at the ends of certain
topological superconductors.~\cite{kitaev2001}

All above mentioned topological phases of matter can be realized as band
insulators or superconductors. Their topological classification goes under the
name of tenfold-way (or $K$-theoretic) classification. The mathematical rules
of the tenfold-way classification state that two given band structures are
topologically equivalent if and only if they can be continuously deformed into
each other without closing the band gap or violating  symmetry constraints. The band
structures with different number of bands can be topologically equivalent too: the tenfold-way classification allows the addition of ``trivial'' bands both
above and below the band gap. 

Initially, the symmetry constraints considered
included time-reversal, particle-hole and sublattice (chiral) symmetries which
led to an elegant classification result containing ten entries with a periodic
structure.~\cite{kitaev2009,schnyder2009} Recently, crystalline symmetries have
been included to extend the tenfold-way classification which now contains many
thousands
entries.~\cite{turner2012,fu2011,trifunovic2017,trifunovic2019,khalaf2018,bradlyn2017,huang2017,shiozaki2014,geier2019,ono2020,khalaf2018b,schindler2018,fu2011,zhang2019,trifunovic2020a}
The extended tenfold-way classification is listed in
catalogues~\cite{bradlyn2017,zhang2019} that helped discovery of  many
topological material candidates. A novel robust effect of some of these topological
crystalline phases are so-called higher-order boundary states: chiral (helical)
modes can appear not only on the boundary of a two-dimensional systems but also
on the hinges of three-dimensional systems. Similarly, Majorana zero-energy
states can appear as corner states of either two- or three-dimensional systems.
These robust boundary effects are guaranteed by the bulk-boundary
correspondence~\cite{schnyder2009,trifunovic2019,trifunovic2020a} that holds for the tenfold-way topological classification.

\begin{figure}[t]
	\centering
	\includegraphics[width=0.8\columnwidth]{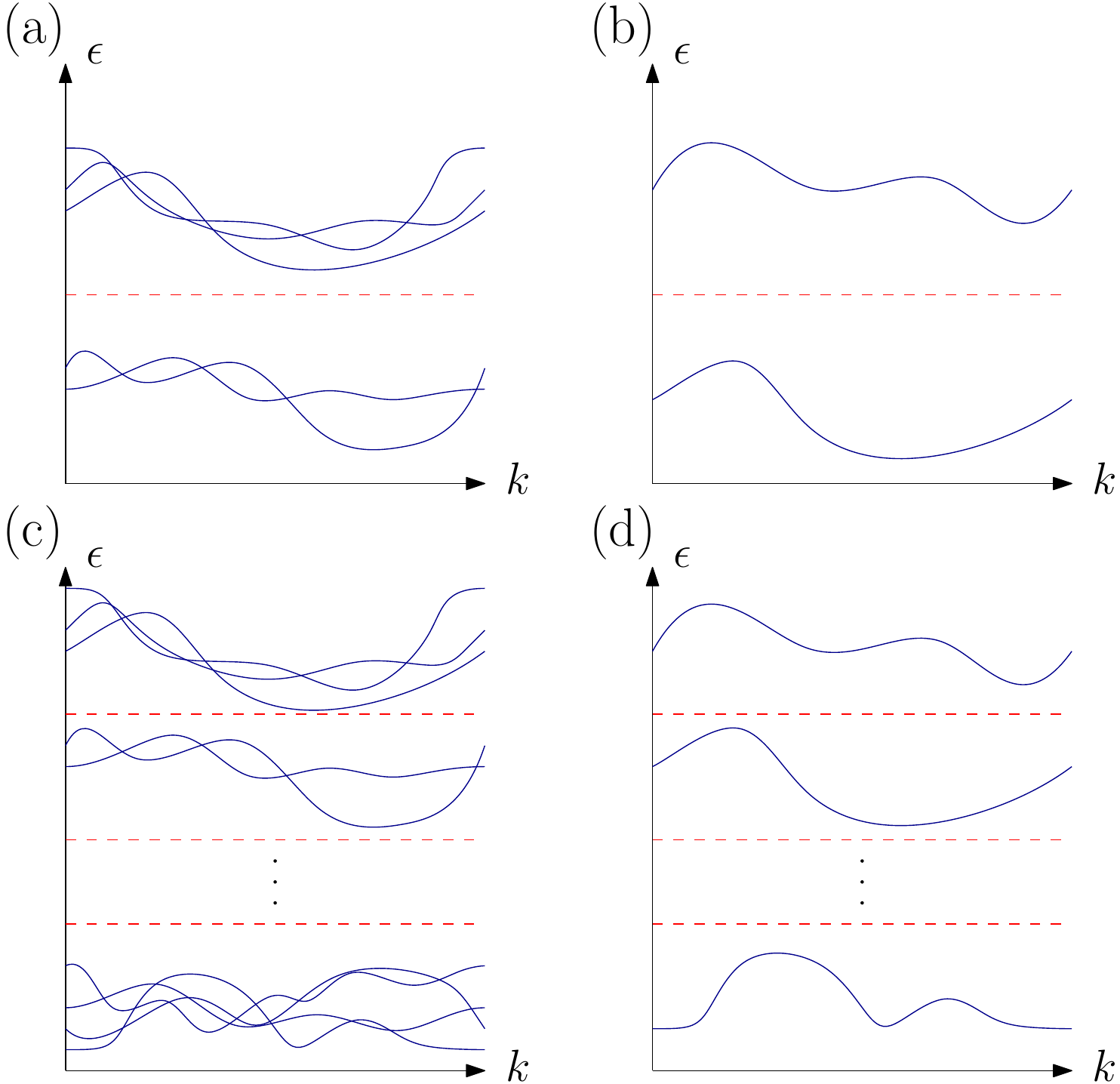}
	\caption{The band structure of a band insulator, where the existence of a single band gap is guaranteed (a). The same as in the panel (a) for the case of two bands (b). The band insulator can have $N-1$ band gaps that divide the bands into $N$ disjoint sets (c). The same as in the panel (c) for the case when each of $N$ sets contains exactly one band (d).
}
	\label{fig:1}
\end{figure}

While the quest for new topological materials is still an ongoing effort, some more
recent theoretical efforts are concerned with the following question: in which
way does a modification of the tenfold-way classification rules alter the established
classification results? Such ``beyond tenfold-way'' classification schemes
include delicate and fragile (i.e., unstable~\footnote{In early studies~\cite{kennedy2014,kennedy2016} both delicate and fragile phases were called unstable topological phases, in order to distinguish them from the stable (tenfold-way) topological phases. In this work we borrow the terminology of Ref.~\onlinecite{nelson2020} and call phases {\it delicate} if they are unstable but not fragile.})  topological classifications. For
delicate classification (Fig.~\ref{fig:1}b), both the number of conduction
and valence bands is fixed. The most well studied representative of delicate topological
insulator is two-band Hopf insulator.~\cite{moore2008,kennedy2014} Recently, many fragile
topological insulators~\cite{po2018,kennedy2014} were accidentally discovered while comparing the
classification results of ``Topological quantum chemistry''~\cite{bradlyn2017}
and that of ``Symmetry-based indicators''.~\cite{po2017} ``Fragile''
topological equivalence allows for the addition of trivial conduction bands while the
number of valence bands is fixed. In other words, the fragile classification
rules are halfway between that of tenfold-way (i.e., stable) and delicate topological classification. Yet another possibility of going beyond the tenfold-way is to introduce additional constraints on the band structure. For example, the boundary-obstructed
classification~\cite{benalcazar2017,khalaf2021} requires that the so-called Wannier gap~\footnote{The Wannier gap,
unlike the band gap, is not physical observable; It is unclear how to define it
for interacting systems.} is maintained.  We note that so far, the efforts were
mainly focused on obtaining such modified classifications.  Despite
efforts~\cite{song2020,alexandradinata2020} to formulate a bulk-boundary
correspondence, it is still unclear if any (possibly subtle) quantized boundary
effect can be used to uniquely identify any of the ``beyond tenfold-way''
phases.

In this work we focus our attention on delicate topological phases. The
constraint that the number of bands needs to be fixed hinders direct
application to crystalline materials. For example, the Hopf insulator has
exactly one conduction and one valence band, whereas crystals have typically
many bands. Although the Hopf insulator can be turned into a stable topological phase through additional symmetry constraints~\cite{liu2017}, here we take a different route and relax the requirements of the delicate topological
classification to allow for a trivial band to be added if
separated by the gaps from all the other bands, see Fig.~\ref{fig:1}d. The idea
of multi-gap classification is not a new one. The stable multi-gap
classification was used to classify Floquet insulators,~\cite{roy2017} whereas
the delicate multi-gap classification, with exception of the one-dimensional
systems described by real Hamiltonians,~\cite{ahn2018,wu2019,apoorv2020} has been largely unexplored. 

\begin{figure}[t]
	\centering
	\includegraphics[width=0.7\columnwidth]{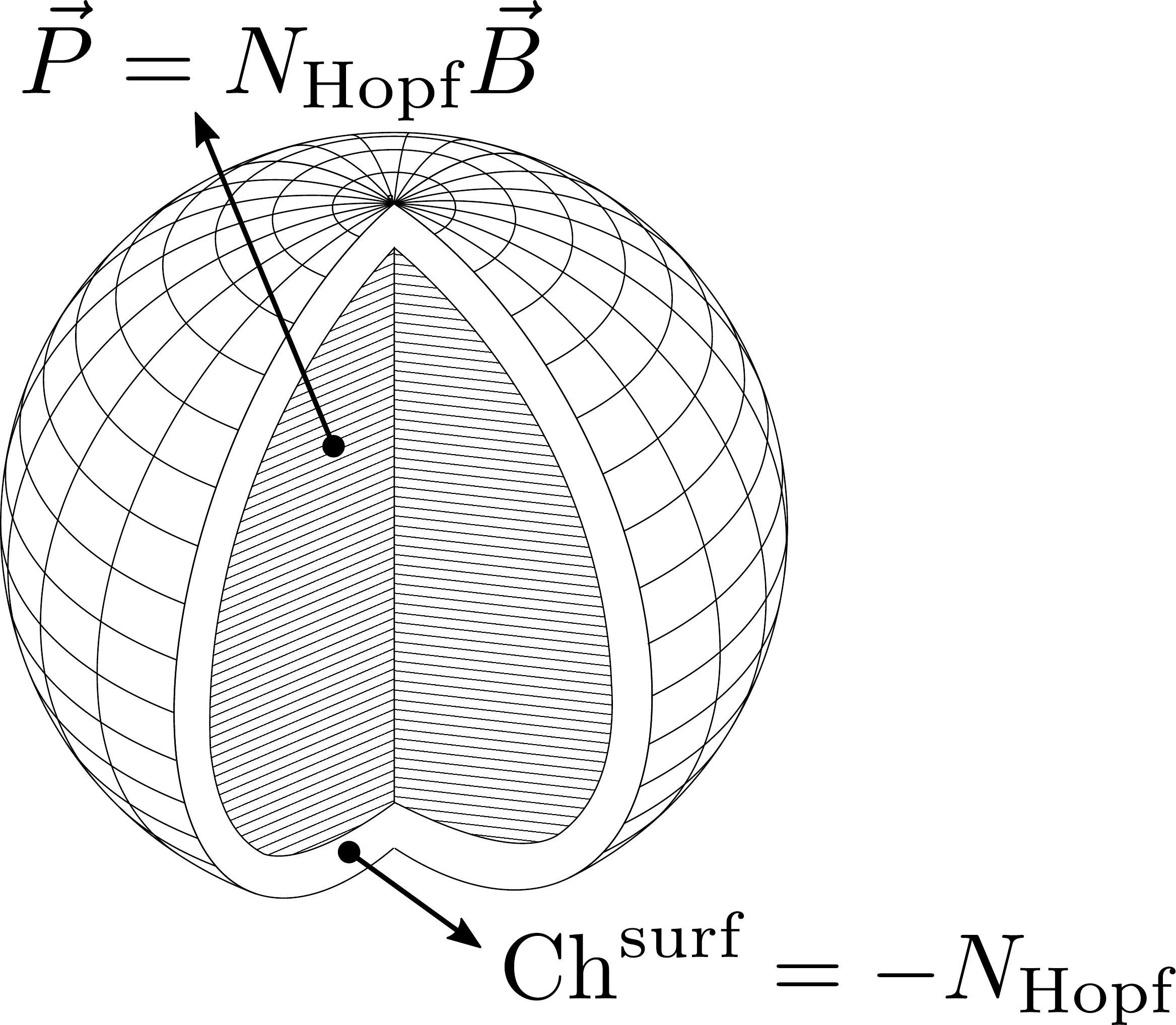}
	\caption{A finite $N$-band Hopf insulator has the bulk with the total (all the bulk bands combined) isotropic magnetoelectric polarizability coefficient equal to $N_\text{Hopf}$ (where $e=h=1$) with the gapped boundary that has Hall conductivity equal to $-N_\text{Hopf}$. The division between the bulk and the boundary is achieved by representing the bulk by set of exponentially localized Wannier functions.
}
	\label{fig:ball}
\end{figure}

We consider three-dimensional systems with no additional symmetry constraints, and
find that delicate multi-gap topological classification is the same as
the classification of the Hopf insulator. The obtained phases are dubbed $N$-band
Hopf insulators. Unlike the tenfold-way topological insulators, the boundary of the $N$-band Hopf insulator can be fully gapped and there is no unique way of defining the boundary subsystem, see Sec.~\ref{sec:bb}. Remarkably, despite this non-uniqueness, we are able to formulate the bulk-boundary correspondence for the
$N$-band Hopf insulator: a finite sample of the $N$-band Hopf
insulator can be seen as the bulk, with the isotropic orbital magnetolectric polarizability coefficient
(of all the bulk bands)~\footnote{The magnetoelectric polarizability tensor is isotropic if the contributions from all the bands are considered, see Ref~\onlinecite{essin2010}.} taking an integer value, wrapped in a Chern insulator sheet with the total Chern number of all the boundary bands equal to minus the same integer, see Fig.~\ref{fig:ball}. Recently, Alexandradinata, Nelson, and Soluyanov~\cite{alexandradinata2020} formulated the bulk-boundary correspondence for $N=2$ Hopf insulator, albeit for a subset of boundary conditions that leave the boundary gapped. In this work we show that such bulk-boundary correspondence is a physical one: the quantized boundary effect can be measured in certain non-equilibrium states. Specifically, a finite sample fully filled with electrons does not exhibit any quantized effect, the quantized response is obtained only by driving the system into
a non-equilibrium state where the region close to the boundary (bulk)  is fully filled with electrons while the bulk (boundary) is unoccupied.

The results of topological classifications apply equally well to periodically
and adiabatically driven crystals. In fact, there is a well known one-to-one
correspondence between a two-dimensional Quantum Hall system and a one-dimensional
Thouless pump.~\cite{thouless1983} The quantized Hall conductance translates
into quantized charge pumped during one period of the adiabatic drive. Analogously,
there is one-to-one correspondence between a three-dimensional $N$-band
topological insulator and certain two-dimensional adiabatic pumps: the quantized magnetoelectric polarizability coefficient of the three-dimensional bulk translates into the quantized orbital magnetization of the two-dimensional bulk of the pump, while the surface Chern number translates into the edge Thouless pump. We
explicitly construct one such two-dimensional $N$-band Hopf
pump which happens to also represents an anomalous Floquet
insulator (AFI).~\cite{rudner2013} Unlike Floquet insulators, where the
condition of the gap in the quasienergy spectrum is difficult to verify
experimentally, the requirements of $N$-band Hopf insulators are
experimentally accessible.

The remaining of the article is organized as follows. In Sec.~\ref{sec:hopf} we
review the definition and the classification of Hopf insulators. Sec.~\ref{sec:Nhopf}
considers $N$-band Hopf insulators and derives their classification and
topological invariant. The bulk-boundary correspondence for $N$-band Hopf
insulators is formulated in Sec.~\ref{sec:bb}. In Sec.~\ref{sec:magnetization},
we consider a two-dimensional $N$-band Hopf pump and discuss its
orbital magnetization. Examples of both three-dimensional Hopf insulator and two-dimensional
$N$-band Hopf pump with $N=2$ and $N=3$ can be found in
Sec.~\ref{sec:examples}. We conclude in Sec.~\ref{sec:conclusions}.

\section{Hopf Insulator}\label{sec:hopf}
Consider a three-dimensional, gapped $2$-band Bloch Hamiltonian $h_\vk$. Assuming
that the two bands are ``flattened'' such that the Bloch eigenvalues become
$\pm1$, we write
\begin{align}
	h_\vk&=U_\vk\sigma_3 U_\vk^\dagger,
	\label{eq:1}
\end{align}
where $\sigma_3$ is Pauli matrix and $U_\vk\in SU(2)$. At each $\vk$-point
in the Brillouin zone (BZ), $h_\vk$ can be seen as an element of the quotient group
$SU(2)/U(1)$, where $U(1)\in SU(2)$ describes gauge transformation that changes the relative phase between the two Bloch eigenvectors. The group $SU(2)/U(1)$ is isomorphic to $2$-sphere $S^2$,
hence $h_\vk$ is seen as a map from BZ (3-torus $T^3$) to $S^2$
$h_\vk:T^3\rightarrow S^2$. This map is defined by the representation of the Bloch state $\ket{u_{\vk 1}}$ on the Bloch sphere. It follows that the
classification of three-dimensional $2$-band Bloch Hamiltonians is given by
homotopy classification of the maps $h_\vk:T^3\rightarrow S^2$. The complete
classification of such maps was first obtained by
Pointryagyn~\cite{pontryagin1941}. There are three weak topological invariants~\footnote{For the tenfold-way classification, the distinction between strong and weak topological invariants is related to the effects of disorder; The value of a strong topological invariant cannot change due to inclusion of translation-symmetry-breaking perturbations. On the other hand, delicate topological phases depend crucially on the presence of translational symmetry. Hence, in this work we use a more general definition, where the strong topological invariants are those invariants that can be defined on a $d$-dimensional sphere instead of BZ.}
classifying the maps from $T^2\rightarrow S^2$ with $T^2\subset T^3$---these
invariants are Chern numbers in $(k_x,k_y)$-, $(k_y,k_z)$- and
$(k_x,k_z)$-manifolds. If any of the weak invariants is non-zero, the homotopy
classification of $h_\vk$ does not have a group structure. In this article we
assume that the weak invariants vanish, in which case the $\ZZ$ classification
is obtained, given by the Hopf invariant $N_\text{Hopf}=2P_3^1$, with $P_3^1$
Abelian the third Chern-Simons form (Abelian axion coupling)
\begin{align}
	P_3^n&=\int_\text{BZ}\frac{d^3 k}{8\pi^2}\vA_n\cdot\vec\nabla\times\vA_n,
	\label{eq:2}
\end{align}
with $\vA_n=i\langle u_{\vk n}\vert\vec\nabla_\vk\vert u_{\vk n}\rangle$.

We now proceed with an alternative derivation of the above results. Vanishing
of the weak invariants implies the homotopy classification of maps $T^3\rightarrow
S^2$ is given by the homotopy group $\pi_3(S^2)$ that classifies maps
$S^3\rightarrow S^2$. Instead of calculating
$\pi_3(SU(2)/U(1))=\pi_3(S^2)$, we calculate the relative homotopy group
$\pi_3(SU(2),U(1))$ which is isomorphic to $\pi_3(SU(2)/U(1))$. The relative homotopy group $\pi_3(SU(2),U(1))$ classifies the maps from the $3$-disc $D^3$ to the group
$SU(2)$ with the constraint that the disc's boundary is mapped to the subgroup
$U(1)$, $\partial D^3\rightarrow U(1)$. The topological invariants for the
group $\pi_3(SU(2),U(1))$ can be obtained from the knowledge of homotopy groups
for $SU(2)$ and $U(1)$ with help of the following exact
sequence~\cite{turner2012,trifunovic2017, hatcher}
\begin{align}
	\label{eq:3}
	\pi_3(U(1))\xrightarrow{i_3}\pi_3(SU(2))&\xrightarrow{i}\pi_3(SU(2),U(1))\\
	&\xrightarrow{\partial}\pi_2(U(1))\xrightarrow{i_2}\pi_2(SU(2)).\nonumber
\end{align}
The exactness of the above sequence means that the image of each homomorphism
is equal to the kernel of the subsequent homomorphism. The homomorphisms $i_3$
and $i_2$ are induced by the inclusion $U(1)\rightarrow SU(2)$, the
homomorphism $i$ identifies the maps from $S^3\rightarrow SU(2)$ as maps
$D^3\rightarrow SU(2)$ where the boundary $\partial D^3$ is mapped to the
identity element of the group $SU(2)$. Lastly, the boundary homomorphism
$\partial$ restricts the map $D^3\rightarrow SU(2)$ to its boundary map
$\partial D^3=S^2\rightarrow U(1)$ which is classified by the group
$\pi_2(U(1))$. In this particular case the groups $\pi_3(U(1))$ and
$\pi_2(U(1))$ are trivial, hence the exactness of the sequence~(\ref{eq:3})
implies
\begin{align}
	\pi_3(SU(2),U(1))=\pi_3(SU(2))=\ZZ.
	\label{eq:4}
\end{align}
The topological invariant for the homotopy group $\pi_3(SU(2))$ is the third winding number $W_3[U_\vk]$, $U_\vk\in SU(2)$
\begin{align}
	W_3[U_\vk]&=\int_\text{BZ}\frac{d^3k}{8\pi^2}\text{Tr}\left( U_\vk^\dagger\partial_{k_x} U_\vk[U_\vk^\dagger\partial_{k_y} U_\vk,U_\vk^\dagger\partial_{k_z}U_\vk]_- \right),
	\label{eq:5}
\end{align}
where $[A,B]_-$ denotes the commutator. Finally, the isomorphism between the groups
$\pi_3(SU(2),U(1))$ and $\pi_3(SU(2)/U(1))$ implies that there is a relation
between the winding number~(\ref{eq:5}) and the Hopf invariant~(\ref{eq:2}).
Indeed, the following relation holds
\begin{align}
	N_\text{Hopf}= W_3[U_\vk]=P_3^1+P_3^2=2P_3^1=N_\text{Hopf},
	\label{eq:6}
\end{align}
where $\vert u_{\vk1}\rangle$ and $\vert u_{\vk2}\rangle$ are two Bloch
eigenvectors that define $P_3^{1,2}$ via Eq.~(\ref{eq:2}), and
$U_\vk:T^3\rightarrow SU(2)$ is defined in Eq.~(\ref{eq:1}). The
relation~(\ref{eq:6}) was proved in Ref.~\onlinecite{unal2019}, we review its derivation in Appendix~\ref{sec:eq6}.

\section{$N$-band Hopf insulators}\label{sec:Nhopf}
The gap of a band insulator divides the Hilbert space into two mutually orthogonal
subspaces, with the projector ${\cal P}_\vk$ (${\cal Q}_\vk\equiv1-{\cal P}_\vk$)
defined by occupied (empty) Bloch eigenvectors; see Fig.~\ref{fig:1}a.
The topological classification of band insulators is obtained by classifying
the subspace ${\cal P}_\vk$, or equivalently ${\cal Q}_{\vk}$. Within the
$K$-theory classification, the ranks of these two projectors, ${\cal P}_\vk$
and ${\cal Q}_\vk$, can be varied by an addition of topologically trivial
bands. On the other hand, the fragile topological classification~\cite{po2018}
allows the ranks of ${\cal Q}_\vk$ to be varied while the rank of the projector
${\cal P}_\vk$ is fixed. If the ranks of both ${\cal P}_\vk$ and ${\cal Q}_\vk$
are required to take some fixed values, as is the case for the $N=2$ band Hopf
insulator in Fig.~\ref{fig:1}b, one then talks about delicate topological
classification.

In this work we modify the classification rules by requiring not one
but $N-1$ band gaps are to be maintained, see Fig.~\ref{fig:1}. Such a band structure
defines $N$ projectors ${\cal P}_{\vk n}$, $n=1,\dots,N$, which are projectors
onto the subspaces spanned by the Bloch eigenvectors with the eigenvalues laying between two neighbouring band gaps.

As in the case of a single band-gap classification, for the $(N-1)$ band-gap
classification one can apply various classification rules. The $K$-theoretic
version of the classification, see Fig.~\ref{fig:1}c, allows the rank of all
projectors ${\cal P}_{\vk n}$ to be varied by the addition of trivial bands---such
classification is directly related to a single band-gap classification, see
Ref.~\onlinecite{roy2017}. On the other hand,~\footnote{There are more
	possibilities herein,~\cite{bouhon2020b} one can define fragile
	classifications by allowing only certain ranks ${\cal P}_\vk^n$ to be
varied, although the physical relevance of such classification schemes is
unclear.} if the rank of all the projectors ${\cal P}_{\vk n}$ is fixed, we
refer to this classification as delicate multi-gap classification. In
contrast to $K$-theoretic classification, the delicate multi-gap classification is not always related to delicate single-gap
classification~\cite{bouhon2020,wu2019}.

Below we show that the delicate $(N-1)$-gap classification of
the three-dimensional Bloch Hamiltonians (with vanishing Chern numbers) is  $\ZZ$  for
$N\ge2$ if $\mathrm{rank}\,{\cal P}_{\vk n}=1$ for $n=1,\dots,N$. Since the
non-trivial topological insulators for $N=2$ are called Hopf
insulators,~\cite{moore2008} we call the non-trivial insulators for $N>2$ 
$N$-band Hopf insulators.

The complete classification of $N$-band Hopf insulators goes along the lines of
$N=2$ classification of Sec.~\ref{sec:hopf}. Given $N$-band Bloch Hamiltonian
$h_\vk$ is flattened such that its eigenvalues are distinct integers $[1,N]$,
 the diagonalized Hamiltonian is written as
\begin{align}
	h_\vk&=U_\vk \,\text{diag}(1,\dots,N)\,U_\vk^\dagger,
	\label{eq:8}
\end{align}
where $U_\vk\in SU(N)$ is continuous on the BZ. At each $\vk$-point in the BZ, $h_\vk$ is seen as an element
of the group $SU(N)/U(1)^{N-1}$, where the subgroup $U(1)^{N-1}\in SU(N)$ is
generated by $U(1)$ gauge transformations of individual bands. Under the
assumption of vanishing weak topological invariants, that are defined for each ${\cal P}_{\vk n}$, the BZ can be regarded as 3-sphere $S^3$.
In other words, the strong classification of $N$-band Hopf insulators is given
by the homotopy group $\pi_3(SU(N)/U(1)^{N-1})$. We proceed with help of the
following isomorphism~\footnote{The group $\pi_i(X,A)$ is not isomorphic to
$\pi_i(X/A)$ in general. For example, when $X=D^2$ and $A=S^1$, the group
$\pi_i(D^2,S^2)$ is trivial for $i>2$ as seen by exact sequence similar to
Eq.~(\ref{eq:3}), whereas the $\pi_i(D^2/S^1=S^2)$ is non-trivial
for infinitely many values of $i$. The isomorphism~(\ref{eq:9}) follows directly from the long exact sequence for the fibration $U(1)^N\rightarrow U(N)\rightarrow U(N)/U(1)^N$.}
\begin{align}
	\pi_3(SU(N)/U(1)^{N-1})&=\pi_3(SU(N),U(1)^{N-1}),
	\label{eq:9}
\end{align}
where $\pi_3(X,A)$ for $A\subseteq X$ denotes the relative homotopy group
introduced in the previous Section. The exact sequence, analogous to the one in Eq.~(\ref{eq:3}), reads
\begin{align}
	\pi_3(U(1)^{N-1})&\xrightarrow{i_3}\pi_3(SU(N))\xrightarrow{i}\pi_3(SU(N),U(1)^{N-1})\nonumber\\
	&\xrightarrow{\partial}\pi_2(U(1)^{N-1})\xrightarrow{i_2}\pi_2(SU(N)),
	\label{eq:10}
\end{align}
implying that $\pi_3(SU(N),U(1)^{N-1})=\pi_3(SU(N))$ because the homotopy groups
$\pi_3(U(1)^{N-1})$ and $\pi_2(U(1)^{N-1})$ are trivial. The topological
invariant, a member of the group $\pi_3(SU(N))=\ZZ$, is the third winding number,
which provides the complete classification of $N$-band Hopf
insulators. As we show in the Appendix~\ref{sec:nonstable1d}, the classification approach used
above can be also applied to the case of real one-dimensional $N$-band systems which
were shown to have non-Abelian classification~\cite{wu2019,bouhon2020}. The
advantage of our classification approach is that it gives the complete set of
topological invariants that were previously not known.

The above considerations give the topological invariant of the $N$-band Hopf insulator
\begin{align}
    N_\text{Hopf}=W_3[U_\vk],
\label{hopfinvarianteq}
\end{align}
i.e., $N_\text{Hopf}$ is the third winding number of the unitary
$N\times N$ matrix $U_\vk\in SU(N)$ in
Eq.~(\ref{eq:8}). Although $U_\vk$ explicitly depends on the choice of $U(1)$
gauge for each Bloch eigenvector, such gauge transformations cannot change the third winding number of $U_\vk$. (This follows directly from the exact
sequence~(\ref{eq:10}), since $\text{img}\, i_3$ is trivial.) The following relation holds
\begin{align}
	N_\text{Hopf}=P_3 \quad \in\,\mathbb{Z},
	\label{eq:12}
\end{align}
where $P_3$ is non-Abelian third Chern-Simons form
\begin{align}
    P_3&=\int_\text{BZ}\frac{d^3 k}{8\pi^2}\mathrm{tr}\left({\vec {\hat A}}_{\vec k}\cdot\vec\nabla\times{\vec {\hat A}}_{\vec k}+\frac{2i}3 {\vec {\hat A}}_{\vec k}\cdot {\vec {\hat A}}_{\vec k}\times {\vec {\hat A}}_{\vec k}\right),
    \label{eq:P3NA}
\end{align}
with $({\vec{\hat A}}_{\vec k})_{nm}=i\langle u_{{\vec k}n}\vert\nabla_{\vec k}\vert u_{{\vec k}m}\rangle$. To prove the relation~\eqref{eq:12}, we note that under a gauge transformation $U_{\vk}$, the non-Abelian third Chern-Simons form transforms in the following way~\cite{ryu2010}
\begin{equation}
P_3 \mapsto \tilde P_3 + W_3[U_{\vec{k}}].
\label{gauge}
\end{equation}
In the basis of orbitals of the unit cell $\{\ket{1},...,\ket{N}\}$, the non-Abelian third Chern-Simons form vanishes, $\tilde P_3=0$. If we apply a gauge transformation $U_{\vec{k}}$, the new basis corresponds to the Bloch eigenvectors $\{\ket{u_{\vk 1}},...,\ket{u_{\vk N}}\}$. In this new basis, by the gauge transformation law \eqref{gauge}, we have that the non-Abelian third Chern-Simons form satisfies $P_3 = W_3[U_{\vec{k}}]$, proving the relation~\eqref{eq:12} using Eq.~\eqref{hopfinvarianteq}.

The above topological invariant differs from the tenfold-way topological invariants, which vanish when summed over all the bands. Furthermore, for tenfold-way classification, the non-Abelian third Chern-Simons form~(\ref{eq:P3NA}) has an integer ambiguity which is removed by requiring $N$ band gaps to stay open, or equivalently, requiring $\ket{u_{\vk n}}$ to be continuous over the BZ for all $n$.

The obtained topological invariant~(\ref{eq:12}) has a physical meaning of the isotropic magnetoelectric polarizability coefficient $\alpha$ of all the bulk bands combined.~\cite{qi2008,essin2010} The magnetoelectric polarizability coefficient is a tensor quantity, which has two contributions:~\cite{essin2010} a topological (isotropic) contribution is given by non-Abelian Chern-Simons form~(\ref{eq:P3NA}) which is equal to $N_\text{Hopf}$ by virtue of Eq.~(\ref{eq:12}), and a non-topological contribution which vanishes in the absence of unoccupied bands. Unlike the tenfold-way topological invariants which can be assigned to each band (or group of bands) separately, the above topological invariant can only be assigned to the whole band structure. Indeed, we can express the isotropic magnetoelectric polarizability coefficient $\alpha$ as
\begin{align}
    N_\text{Hopf}=\alpha=\sum_{n=1}^N\alpha_n,
\end{align}
where $\alpha_n$ is the isotropic component of the magnetoelectric polarizability tensor for the $n$th band. There are two contributions~\cite{essin2010} to the magnetoelectric polarizability coefficient $\alpha_n=\alpha_n^\text{top}+\alpha_n^\text{nontop}$, where the topological piece $\alpha_n^\text{top}$ is expressed via the Abelian third Chern-Simons form~(\ref{eq:2}) that involves only Bloch eigenvector of the $n$th band
\begin{align}
    \alpha_n^\text{top}=P_3^n,
\end{align}
while for the non-topological piece $\alpha_n^\text{nontop}$, the knowledge of the whole band structure is required.~\cite{essin2010} We note that, generally, a non-quantized value of $\sum_{n=1}^N\alpha_n^\text{top}$ ($\sum_{n=1}^N\alpha_n^\text{nontop}$) cannot change upon a deformation of the Hamiltonian that maintains all $N-1$ gaps.

\section{Bulk-boundary correspondence}\label{sec:bb}
\begin{figure}[t]
	\centering
	\includegraphics[width=.8\columnwidth]{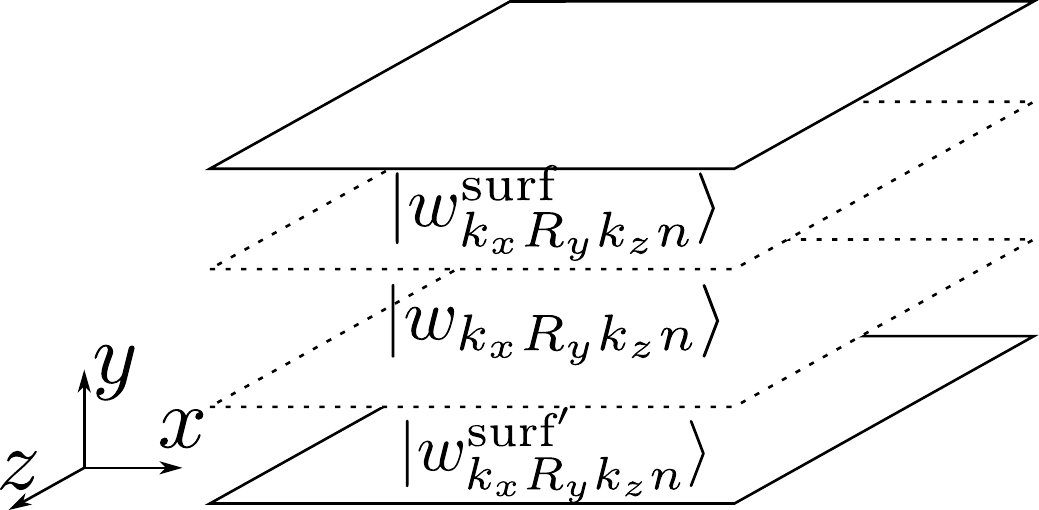}
	\caption{The Hilbert space, spanned by the orbitals of the slab's
		supercell, is divided at each $(k_x,k_z)$-point into the three
		mutually orthogonal subspaces corresponding to the bulk and the
		two surfaces. The bulk hybrid WFs $\ket{w_{k_xR_yk_zn}}$ are
		continuous in the $(k_x,k_z)$-space. On the other hand, for a
		non-trivial $N$-band Hopf insulator, there is an obstruction in
		finding continuous surface WFs $\ket{w_{k_xR_yk_zn}^\text{surf}}$.}
	\label{fig:3Wcut}
\end{figure}

To formulate bulk-boundary correspondence for $N$-band Hopf insulator, we
consider slab geometry with arbitrary termination along $y$-direction described
by the $NN_y\times NN_y$ slab Hamiltonian $h_{k_xk_z}$. We assume that all
weak topological invariants (Chern numbers) vanish, hence, there exist continuous bulk Bloch
eigenfunctions $\vert\psi_{\vk n}\rangle$, $n=1,\dots,N$, of the Hamiltonian
$h_{k_xk_z}$. In other words, each bulk band can be separately ``Wannierized'': the many-body wavefunction of the fully occupied $n$th band can be obtained by occupying exponentially localized single-electron bulk Wannier functions (WFs) $\ket{w_{\vec R n}}$. The bulk WFs are obtained from continuous Bloch eigenfunctions
\begin{align}
	\vert w_{\vR n}\rangle=\frac{1}{\sqrt{N_xN_yN_z}}\sum_\vk e^{i\vk\cdot\vR}\vert\psi_{\vk n}\rangle.
	\label{eq:wRn}
\end{align}
For a slab terminated in $y$-direction, we use hybrid bulk WFs
\begin{align}
	\vert w_{k_xR_yk_z n}\rangle&=\frac{1}{\sqrt{N_xN_z}}\sum_{R_x,R_z} e^{-i(k_x R_x+k_z R_z)}\vert w_{\vR n}\rangle.
	\label{eq:wzn}
\end{align}
The goal is to divide the slab into the three subsystems: the two surfaces and the bulk, the latter being defined by the choice of the bulk WFs, see Fig.~\ref{fig:3Wcut}.
Using the above WFs we perform a Wannier cut~\cite{trifunovic2020} on all
the bands~\footnote{Since the Wannier cut is performed on all the bands, unlike in
Ref.~\onlinecite{trifunovic2020}, no condition on the crystal's termination
needs to be imposed, i.e., a metallic termination is allowed.} to obtain the projector ${\cal P}_{k_xk_z}^L$ onto the two surfaces by removing the hybrid bulk WFs from the middle of the slab
\begin{align}
	{\cal P}_{k_xk_z}^L(\vx^\prime,\vx)=&\delta_{\vx^\prime\vx}-\sum_{\substack{n=1\\ R_y=-L}}^{\substack{n=N\\ R_y=L}}w_{k_xR_yk_zn}(\vx^\prime)^*w_{k_xR_yk_zn}(\vx),
	\label{eq:PL}
\end{align}
which, for large enough integers $N_y$, $L$ with $N_y\gg (N_y-2 L)$ and $2L<N_y$, defines the projector onto the upper surface
\begin{align}
	{\cal P}^\text{surf}_{k_xk_z}(\vx^\prime,\vx)\equiv&{\cal P}_{k_xk_z}^L(\vx^\prime,\vx) \theta(y)\theta(y^\prime),
	\label{eq:Psurf}
\end{align}
where $\vx=(x,y,z)$ indexes the orbitals of the slab supercell, $\theta(y)$ is the Heaviside theta function, and we assume that the $y=0$ plane passes through the middle of
the slab. The integer $L$ should be chosen as large as possible while requiring
that in the region where the bulk WFs $\vert w_{k_xLk_zn}\rangle$ have support, the
Hamiltonian $h_{k_xk_z}$ is bulk-like.  For the slab's width much larger than the
WFs' size, the operator ${\cal P}^\text{surf}_{k_xk_z}$ is a projector. In fact, thanks to exponential localization of the bulk WFs, ${(\cal P}^\text{surf}_{k_xk_z})^2-{\cal P}^\text{surf}_{k_xk_z}$ converges exponentially to 0 as the slab's width is increased. Hence, the first Chern number of ${\cal P}^\text{surf}_{k_xk_z}$, denoted by
$\text{Ch}^\text{surf}$, reads
\begin{align}
	\text{Ch}^\text{surf}=&i\int_\text{BZ} \frac{dk_xdk_z}{2\pi}\text{Tr}\left( {\cal
	P}^{\text{surf}}_{k_xk_z}[\partial_{k_x}{\cal P}^{\text{surf}}_{k_xk_z},\partial_{k_z} {\cal P}^{\text{surf}}_{k_xk_z}]_-\right).
	\label{eq:C1Psurf}
\end{align}
The bulk-boundary correspondence states
\begin{align}
	\text{Ch}^\text{surf}=-N_\text{Hopf}.
	\label{eq:bb}
\end{align}
The above correspondence can be proved by noticing that $\text{Ch}^\text{surf}$ cannot be changed by surface decorations
since their first Chern number summed over all the bands vanishes. In the
previous section we proved that $N_\text{Hopf}$ is the unique bulk topological
invariant of the $N$-band Hopf insulator, it follows that $\text{Ch}^\text{surf}$
can be expressed in terms of $N_\text{Hopf}$. Hence, to prove the
relation~(\ref{eq:bb}) it is sufficient to show that it holds for the generators
of the $N$-band Hopf insulator, see Sec.~\ref{sec:examples}. We note that compared to tenfold-way classification, where the topological classification group structure is given by the direct sum of two Hamiltonians, the group structure of the classification of the $N$-band Hopf insulator is obtained by concatenation of the BZs of the two band structures. Hence, whereas for tenfold-way topological phases there is a single generator for the classification group $\ZZ$, for the $N$-band Hopf insulator there is one generator for each $N$. Alternatively, the relation~(\ref{eq:bb}) follows from Eq.~(\ref{eq:12}) and ``Surface theorem for axion coupling'' of Ref.~\onlinecite{olsen2017}. The correspondence~(\ref{eq:bb}), for $N=2$, is a generalization of recently discussed bulk-boundary correspondence for the Hopf insulator.~\cite{alexandradinata2020,zhu2021}

The above procedure divides a finite sample of the $N$-band Hopf insulator into
bulk and surface subsystems. It is important to note that such division is not
unique. Choosing different bulk WFs or different assignment of the bulk WFs
to their home unit cell yields different bulk and surface subsystems.\footnote{A direct consequence of this non-uniqueness is inability to uniquely define edge polarization and quadrupole moment of two-dimensional insulators~\cite{trifunovic2020,ren2021}} Despite
this non-uniqueness, a finite sample of the $N$-band Hopf insulator can be seen
to consist of the bulk, with isotropic magnetoelectric polarizability coefficient~\cite{essin2010} being quantized to $\alpha=N_\text{Hopf}$, ``wrapped'' into a sheet
of a Chern insulator with the total Chern number being equal to
$-N_{\text{Hopf}}$, see Fig.~\ref{fig:ball}. (To define the Chern number one considers torus geometry of the boundary.) Clearly, such a ``wrapping paper'' cannot exist as a
standalone object since the total Chern number (of all the bands) of a
two-dimensional system needs to vanish.

Recently,~\cite{nelson2020} the concept of multicellularity for band insulators was discussed. A band insulator is said to be multicellular if it can be Wannierized and if it is not possible to deform the band structure such that all the bulk WFs are localized within a single unit cell. The examples of multicellular band structures include the $N=2$ Hopf insulator and certain insulators constrained by crystalline symmetries. The bulk-boundary correspondence~(\ref{eq:bb}) implies that the $N$-band Hopf insulator is a multicellular phase: if all the bulk WFs are to be localized within a single unit cell, the resulting projector onto the upper surface~(\ref{eq:Psurf}) would be $(k_x,k_z)$-independent and the surface Chern number~(\ref{eq:C1Psurf}) would vanish.

If a finite $N$-band Hopf insulator, fully filled with electrons, is placed into an
external magnetic field, the bulk gets polarized due to the isotropic magnetoelectric effect,
$\vec P=\alpha\vec B=N_\text{Hopf}\vec B$. This polarization does not result in
an excess charge density at the boundary, because the excess charge is
compensated by the surface Chern insulator, which is a direct consequence of the Streda formula~\cite{streda1982} when applied to the surface subsystem. Hence, we see that the two quantized
effects, one in the bulk and the other on the boundary, mutually cancel. It is easy to understand this cancellation by noticing that the many-body wavefunction of a fully occupied slab is independent of the Hamiltonian. Therefore, the fully occupied slab exhibits no magnetoelectric effect, implying that the bulk and the boundary magnetoelectric effects mutually cancel. In order to measure a quantized effect, one needs to drive the system into a non-equilibrium state where either the boundary or the bulk subsystem are fully filled with electrons, but not both.

\section{Orbital magnetization}\label{sec:magnetization}
Every three-dimensional Bloch Hamiltonian $h_\vk$ of a band insulator defines a periodic adiabatic pump of a two-dimensional band structure, and vice versa.
The substitution $k_z\rightarrow2\pi t/T$ gives the Hamiltonian of the
two-dimensional adiabatic pump $h_{k_xk_yt}$ corresponding to the
three-dimensional Hamiltonian $h_\vk$.  As we discuss below, this viewpoint
sheds light on the link between the $N$-band Hopf insulators, introduced in
this work, and the recently studied anomalous Floquet insulator;~\cite{rudner2013}
see Appendix~\ref{sec:floquet} for comparison between $N$-band Hopf pumps and Floquet
insulators.

\begin{figure}[t]
	\centering
	\includegraphics[width=\columnwidth]{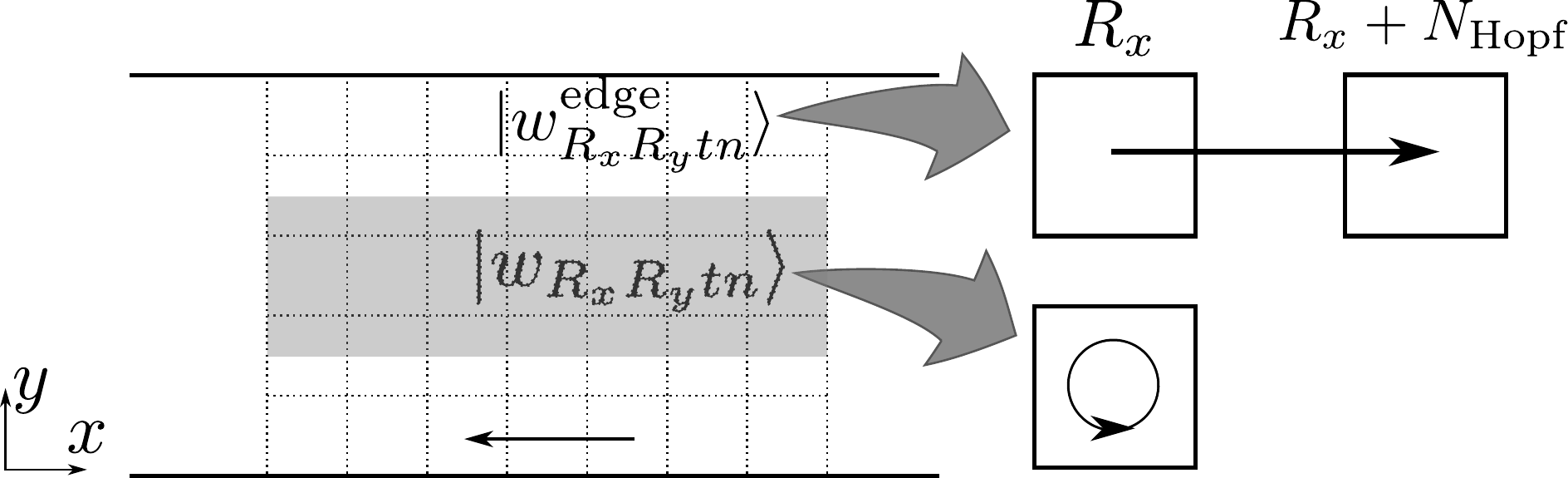}
	\caption{The adiabatic process of a ribbon corresponding to $N$-band
	Hopf pump. The ribbon's supercell is divided into the three
regions, where the middle (gray) region is spanned by the bulk WFs. Whereas the bulk
WFs perform periodic motion, some WFs of the edge subsystems get shifted to the left or right.}
	\label{fig:2Wcut}
\end{figure}

We start by applying the bulk-boundary correspondence~(\ref{eq:bb}) to the
$N$-band Hopf pump $h_{k_xk_yt}$. Consider a ribbon
$h_{k_xt}^\text{ribb}$ consisting of $N_y$ unit cells in $y$-direction. Similar
to Eq.~(\ref{eq:Psurf}), we divide the ribbon-supercell Hilbert space into the
two edge and the bulk subspaces
\begin{align}
	\mathbbm{1}_{NN_y\times NN_y}&={\cal P}^\text{edge}_{k_xt}+{\cal P}^\text{bulk}_{k_xt}+{\cal P}^{\text{edge}\prime}_{k_xt},
	\label{eq:split}
\end{align}
where the right-hand side is the sum of three mutually orthogonal projectors.
Importantly, the ${\cal P}^\text{bulk}_{k_xt}$ projects onto the space spanned
by the bulk hybrid WFs $\vert w_{k_x R_ytn}\rangle$ with $R_y\in [-L,L]$, and the spaces onto which
${\cal P}^\text{edge}_{k_xt}$ and ${\cal P}^{\text{edge}\prime}_{k_xt}$ project
do not contain the orbitals from the middle of the ribbon. This way, at each
$(k_x,t)$-point the ribbon is divided into the bulk and the two edge
subsystems, see Fig.~\ref{fig:2Wcut}. The WFs in the bulk subsystem can be
chosen to be periodic,
\begin{align}
	\vert w_{k_x R_yTn}\rangle&=\vert w_{k_x R_y0n}\rangle,
	\label{eq}
\end{align}
i.e., the bulk WFs return to their initial state after one period. On the other
hand, from the bulk-boundary correspondence~(\ref{eq:bb}), it follows that the upper edge ${\cal
P}_{k_xt}^\text{edge}$ has non-zero Chern number equal to $N_\text{Hopf}$. As a
consequence, the edge WF $\vert w_{k_x R_y0m}^\text{edge}\rangle$ is shifted to
$\vert w_{k_x R_y+N_\text{Hopf};Tm}^\text{edge}\rangle$ for some
$m\in[1,N]$.~\footnote{The more precise statement is that the total shift
	from all the edge bands is $N_\text{Hopf}$, i.e., the shift does not
need to be carried by a single band.} The edge subsystem acts as a Thouless pump
even after considering all the bands---such a situation cannot occur for a
standalone one-dimensional system.

Let us consider a fully occupied ribbon. From the relation~(\ref{eq:12}) and
the results of Ref.~\onlinecite{trifunovic2019a}, we have that the bulk
subsystem has (geometric) orbital magnetization equal to $eN_\text{Hopf}/T$.
To see this, we consider all contributions to orbital magnetization~\cite{thonhauser2005,trifunovic2019a}
\begin{equation}
  \mathfrak{m}= \mathfrak{m}_{\text{pers}}+\mathfrak{m}^{\text{top}}+\mathfrak{m}^{\text{non-top}},
\end{equation}
where the last two terms are the topological and non-topological contribution to the geometric orbital magnetization, and the first term represents the contribution from persistent currents that may exist in the absence of adiabatic drive. Using the relation $\mathfrak{m}^{\text{top}}T=P_3$, we conclude that the topological contribution to orbital magnetization is quantized and equal to $N_\text{Hopf}/T$, see Eq.~\eqref{eq:12}. On the other hand, $\mathfrak{m}^{\text{non-top}}=0$ when all the bands are occupied. Finally, the contribution from persistent currents has to vanish for a fully filled system: such contribution is given by the change of the total energy $E_\text{tot}$ of the system induced by external magnetic field $B$ perpendicular to the system, $\mathfrak{m}_{\text{pers}}=-\frac{\partial E_\text{tot}}{\partial B}$. It follows that $\mathfrak{m}_\text{pers}$ vanishes because $E_{\text{tot}}=\text{Tr}(H_{B})=\text{Tr}(H_{B=0})$, since the external magnetic field only enters in non-diagonal components (in the position basis) of the Hamiltonian. Therefore, we conclude that the orbital magnetization is quantized and given by the Hopf invariant. The orbital magnetization gives rise to an edge current that exactly cancels the current pumped by the edge subsystem. Hence, the bulk and the boundary anomalies mutually cancel similar to the three-dimensional case discussed at the end of the previous Section.

\begin{figure}[t]
	\centering
	\includegraphics[width=1\columnwidth]{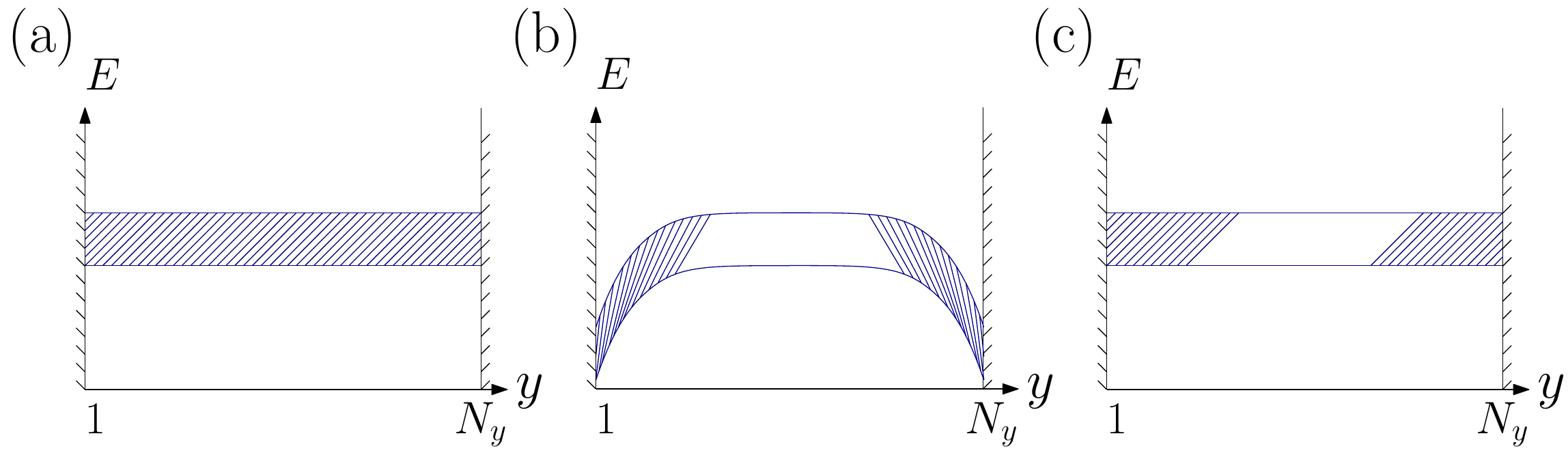}
	\caption{(a) Energy bands of a ribbon, finite in $y$-direction, as a function of $y$-coordinate, where all the states are fully filled with electrons. (b) After applying a gate voltage an equilibrium state is reached where the electrons from the middle of the ribbon are empty (i.e., they move to the drain gate). (c) After switching-off the gate voltage, a non-equilibrium state is obtained with the states close to the edges filled.}
	\label{fig:drivenband}
\end{figure}

In order to observe the quantized orbital magnetization, we need to prepare the ribbon at time $t=0$ such that only the regions close to the edges are fully filled with electrons. To achieve such an initial state, we start from the fully filled band structure illustrated Fig.~\ref{fig:drivenband}(a), and apply a gate voltage, such that in equilibrium, the states in the middle of the ribbon are emptied, as illustrated in Fig.~\ref{fig:drivenband}(b). After the gate voltage is switched-off, the desired initial non-equilibrium state is obtained, as shown on Fig.~\ref{fig:drivenband}(c).  Such an initial state will generally diffuse under the time evolution and eventually electrons leak into the bulk, in which case, as discussed above, no quantization of the orbital magnetization is expected.~\footnote{Such
``leakage'' occurs also for time-independent band insulators””~\cite{ren2021}} Hence, the quantized (geometric) orbital magnetization can be measured in the transient state where the filled regions are separated by an empty bulk. The flat band limit, see Sec.~\ref{sec:examples}, is a special case where the diffusion coefficient is fine-tuned to zero.

The above conclusions parallel the discussion of the so-called anomalous
Floquet insulator (AFI).~\cite{rudner2013} This
is not a coincidence, since in Sec.~\ref{sec:examples}, we show that the $N$-band Hopf pump can, at the same timem be an AFI, although not every AFI is a $N$-band Hopf insulator nor vice-versa. For comparison, in Appendix~\ref{sec:floquet}, we review the stable multi-gap classification of two-dimensional Floquet insulators. One important difference between Floquet insulator and $N$-band Hopf pump is that the latter is not stable against translation-symmetry-breaking perturbations. Indeed, as we discuss in Appendix~\ref{app:2UC}, doubling of the unit cell violates the condition of having a single band between the two neighbouring band gaps.

\section{Examples}\label{sec:examples}
Below, we first consider the three-dimensional Moore-Ran-Wen
model~\cite{moore2008} ($N=2$ band Hopf insulator), that we use to illustrate the bulk-boundary correspondence of Sec.~\ref{sec:bb}, which generalizes the approach of Ref.~\onlinecite{alexandradinata2020}. Furthermore, two two-dimensional examples corresponding to periodic adiabatic processes are considered, which clarify the relation between the $N$-band Hopf insulator and the AFI.

\subsection{Moore-Ran-Wen model of Hopf insulator}
Here we present an example of a 2-band three-dimensional Hopf insulator, the Moore-Ran-Wen model. The Bloch Hamiltonian is defined as~\cite{moore2008}
\begin{equation}
h_{k_xk_yk_z}=\vec{v}\cdot\vec{\sigma},
\label{mrwmodel}
\end{equation}
with $v_i=\vec{z}^{\dagger}\sigma_i\vec{z}$, where $\vec{z}=(z_1,z_2)^T$, with $z_1=\sin(k_x)+i\sin(k_y)$ and $z_2=\sin(k_z)+i[\cos(k_x)+\cos(k_y)+\cos(k_z)-\frac{3}{2}]$. The above model~(\ref{mrwmodel}) has $N_\text{Hopf}=1$. In the following we apply the procedure described in Sec.~\ref{sec:bb} to obtain the surface Chern number \eqref{eq:C1Psurf} for a three-dimensional lattice with $N_x\times N_y\times N_z$ unit cells. The two normalized eigenvectors of the Bloch Hamiltonian \eqref{mrwmodel} are
\begin{align}
    \ket{u_{\vec{k}1}}&=\vert\vec{z}\vert^{-1}(z_1,z_2)^T,\nonumber\\
    \ket{u_{\vec{k}2}}&=\vert\vec{z}\vert^{-1}(z_2^*,-z_1^*)^T,
\end{align}
which are continuous functions of $\vec{k}$. We extend these two Bloch eigenvectors to the whole lattice by defining $\psi_{\vec{k}n}(\vx)=e^{-i\vec{k}\cdot\vx}u_{\vec{k}n}(\vx)$.
\begin{figure}[t]
	\centering
	\includegraphics[width=\columnwidth]{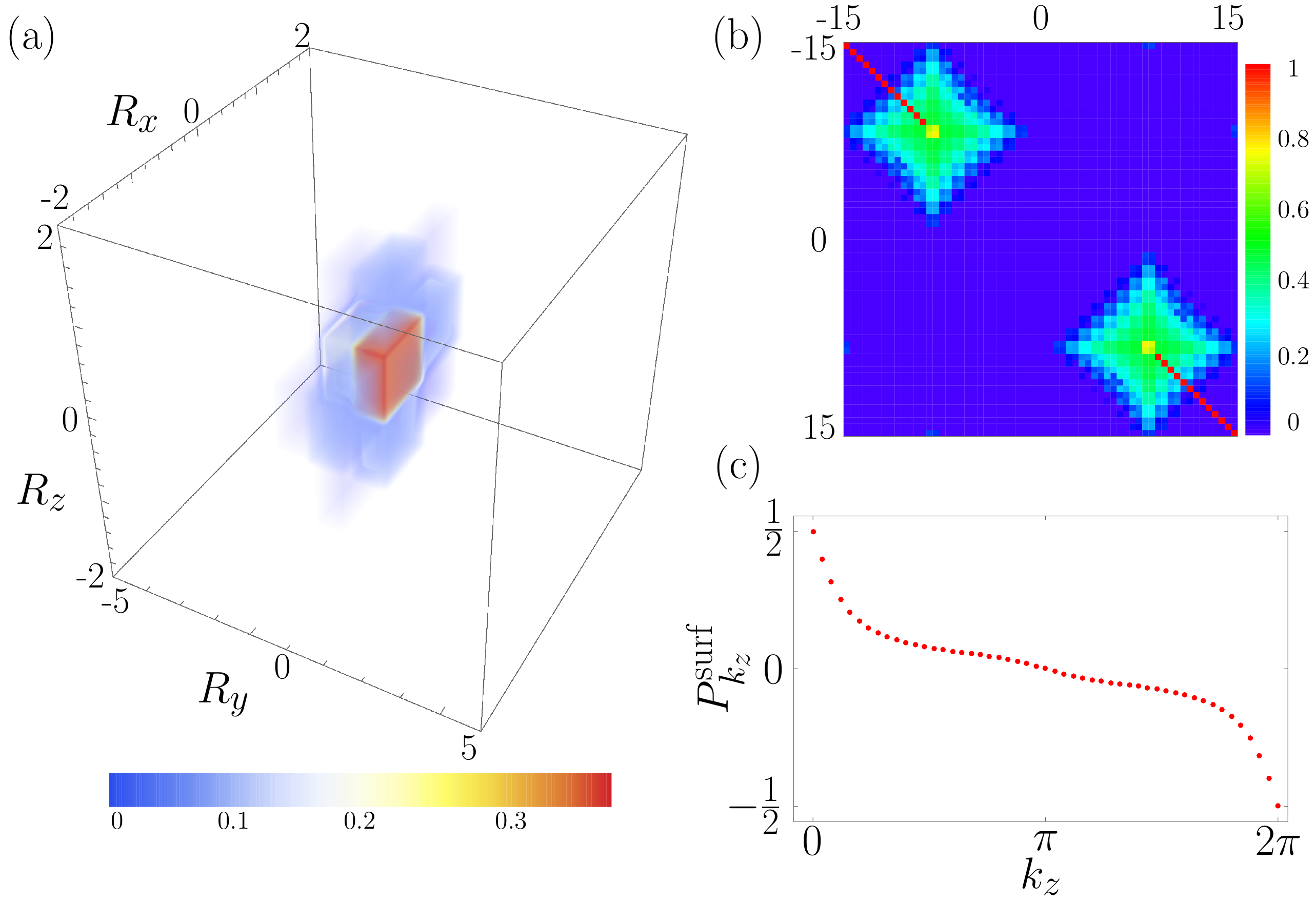}
	\caption{(a) The charge density $\vert w_{(0,0,0)1}(\vx)\vert^2$ of the Wannier function $\ket{w_{(0,0,0)1}}$ localized at the center of the lattice, for $N_y=11$, $N_x=N_z=5$. (b) The absolute value of the matrix elements of the projector in Eq.~(\ref{eq:PL}), $\vert\mathcal{P}_{k_xk_z}^{L}(\vx,\vx^\prime)\vert$, for $N_y=31$ with cutoff $L=8$. (c) The surface polarization along $x$-direction $P_{k_z}^\text{surf}$ for $k_z$ between 0 and $2\pi$ for $N_y=31$ with cutoff $L=8$.}
	\label{fig:MRWplots}
\end{figure}

The WFs $\ket{w_{(0,0,0)n}}$, $n=1,2$, with the home unit cell at $\vec{R}=(0,0,0)$ are given by Eq.~(\ref{eq:wRn}) and shown in Fig.~\ref{fig:MRWplots}a. For arbitrary $\vec{R}=(x,y,z)^T$, the WFs are obtained from the components $w_{(0,0,0)n}(\vx)$ of $\ket{w_{(0,0,0)n}}$
\begin{equation}
w_{\vec{R}n}(\vec x)=w_{(0,0,0)n}(\vx-\vR).
\end{equation}
We use the above choice of the bulk WFs to define the bulk subsystem. To this end, we perform the Fourier transform in $x$- and $y$-directions to obtain the hybrid bulk WFs $\ket{w_{k_xR_yk_zn}}$. Considering only the components $w_{k_xR_yk_zn}(\vx)$ of the hybrid bulk WFs with $\vx$ in a supercell, we obtain the $2N_y\times 2N_y$ projector $\ket{w_{k_xR_yk_zn}}\bra{w_{k_xR_yk_zn}}$. From Eq.~\eqref{eq:PL} we compute the projector $\mathcal{P}_{k_xk_z}^{L}$ onto the two surfaces. As shown in Fig.~\ref{fig:MRWplots}b, after removing the hybrid bulk WFs assigned to the units cells at $R_y\in [-8,8]$, the two surfaces do not overlap and ${\cal P}_{k_xk_z}^\text{surf}$ is obtained from the upper-left block of the matrix $\mathcal{P}_{k_xk_z}^{L}$. The surface Chern number can be obtained
from the $k_z$-dependent surface polarization of all the bands
\begin{align}
    P_{k_z}^\text{surf}=-\frac{i}{2\pi}\ln{{\det}^\prime} \prod_{k_x}\mathcal{P}^{\text{surf}}_{k_xk_z},
\end{align}
where $\det^\prime(X)$ denotes the product of the non-zero eigenvalues of the matrix $X$. The surface Chern number $\text{Ch}^\text{surf}$ implies that the surface polarization $P_{k_z}^\text{surf}$ is not continuous as functions of $k_z$ but jumps by $\text{Ch}^\text{surf}$. The winding of $P_{k_z}^\text{surf}$ is shown in Fig.~\ref{fig:MRWplots}c, where the surface polarization winds once, implying that $\text{Ch}^{\text{surf}}=-1$.

\subsection{Two-dimensional Hopf pumps}
Here we present examples of $N=2$ and $N=3$ Hopf pumps. The
adiabatic evolution $h_{k_xk_yt}$ is piecewise defined, where each
time-segment describes an adiabatic transfer of an electron between  two
selected orbitals of  the  two-dimensional square lattice.

\begin{figure}[t]
	\centering
	\includegraphics[width=\columnwidth]{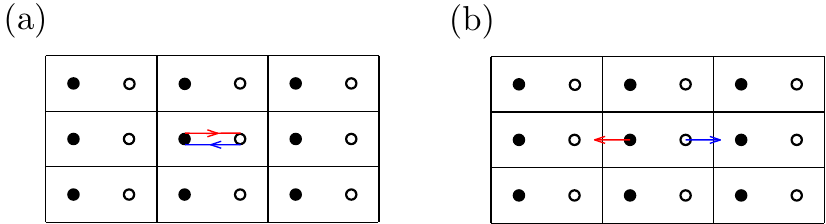}
	\caption{Two different adiabatic processes. The full and empty dots denote $\ket{\vR1}$ and $\ket{\vR2}$ orbitals. If the orbital $\ket{\vR1}$ is occupied, the electron is
		adiabatically transferred (red arrow) to the orbital $\ket{\vR2}$, see Eq.~(\ref{eq:20}). At the end of such process the orbital $\ket{\vR1}$ is	empty while the orbital $\ket{\vR2}$ is occupied. The very same adiabatic process (a)
transfers the electron from the orbital $\ket{\vR2}$ to the orbital $\ket{\vR1}$ as indicated by blue arrow. The second adiabatic process corresponds to an incompete transfer between the two orbitals (b). In that case the final occupied states is
superposition orbitals $\ket{\vR1}$ and $\ket{\vR2}$.}
\label{fig:2}
\end{figure}

The models considered in this subsection are most easily specified pictorially.
In Fig.~\ref{fig:2} we consider adiabatic process in a
system with $2$-sites per unit cell. At $t=0$, the orbitals $\ket{\vR1}$ (black dots) have negative energy whereas the orbitals $\ket{\vR2}$ (empty dots) have positive energy (see Fig.~\ref{fig:2}). We consider the following ``building-block''
adiabatic process
\begin{align}
	h_t&=Be^{-i\sigma_2\pi t/T}\sigma_3e^{i\sigma_2\pi t/T},
	\label{eq:20}
\end{align}
where the Pauli matrices act in the space spanned by the two orbitals.
For the initial state $\ket{\vR1}$, the adiabatic process is
depicted in Fig.~\ref{fig:2}a by the red arrow. The evolution of the excited state
$\ket{\vR2}$ is shown in Fig.~\ref{fig:2}a with the blue arrow. Lastly, one
can stop the above adiabatic process at times $t<T$,
in which case the charge transfer between the sites $\ket{\vR1}$ and $\ket{\vR2}$ is
incomplete. The final state is then a superposition of $\ket{\vR 1}$ and
$\ket{\vR 2}$, as shown in Fig~\ref{fig:2}b. The pictorial representation of the adiabatic process consists of oriented line segments. The start (end) point of a line segment corresponds to the initial (final) state. Below we consider the adiabatic processes where the end points of the line segments lie either on the lattice sites or on the line segment connecting the two neighbouring lattice sites. In the latter case, the initial (final) state is the superposition of the two orbitals located at these two neighbouring sites.
In the following, the number of
arrows enumerates the time-segments, for example, ``$\rightarrow$'' describes
the first segment, ``$\twoheadrightarrow$'' the second etc. The two examples
that follow consider translationally invariant systems, hence the adiabatic
process~(\ref{eq:20}) is extended in a translationally-symmetric manner to
the whole two-dimensional lattice. 

\subsubsection{$N=2$ band Hopf pump}\label{sec:N2Hopfp}
\begin{figure}[t]
	\centering
	\includegraphics[width=\columnwidth]{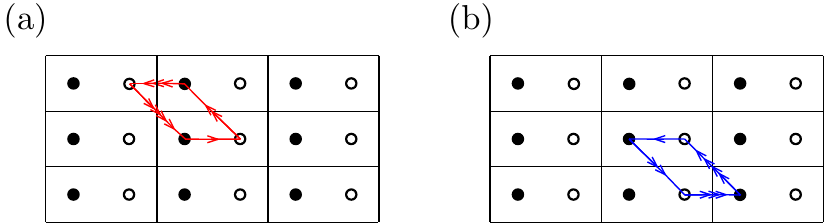}
	\caption{Two level periodic drive made of 4 steps of equal duration.
		Panel (a) shows adiabatic evolution for the initial states
		$\ket{1}$. The same as the panel (a) for the initial
		state $\ket{2}$ (b).}
	\label{fig:3}
\end{figure}

Here, we consider a periodically driven system with two states
per unit cell, labeled by $\{\ket{\vR1},\ket{\vR2}\}$, where
the vector $\vR$ belongs to the square two-dimensional lattice. The driving
protocol is of period $T$ and is made of 4 steps of equal duration
$\frac{T}{4}$, see Fig.~\ref{fig:3}.
At each of those steps, the Hamiltonian reads
\begin{align}
	h_{k_xk_yt}&=\mathcal{U}^{\dagger}_{k_xk_yt} e^{-2\pi i \sigma_2t/T}B\sigma_3e^{2\pi i
	\sigma_2t/T}\mathcal{U}_{k_xk_yt},
	\label{eq:hN2}
\end{align}
with 
\begin{equation}
	\mathcal{U}_{k_xk_yt}=\begin{cases}\sigma_0 & \text{$t\in[0,\frac{T}{4})$},\\
\text{diag}(1,e^{-ik_y}) & \text{$t\in[\frac{T}{4},\frac{T}{2})$},\\
\text{diag}(e^{ik_x},1) &\text{$t\in[\frac{T}{2},\frac{3T}{4})$},\\
\text{diag}(1,e^{-i(k_x-k_y)}) &\text{$t\in[\frac{3T}{4},T)$},
\end{cases}
\label{eq:S4}
\end{equation}
where the Pauli matrices $\sigma_i$ act on the space spanned by the two orbitals in
the unit cell. This two-band Hamiltonian can equivalently be written as the
Hamiltonian of a spin in a time- and momentum-dependent magnetic field,
$h_{k_xk_yt}=\vec{B}_{k_xk_y t}\cdot \vec{\sigma}$. Therefore, the unitary transformation  in Eq.~(\ref{eq:1}) is given by
\begin{align}
	U_{k_xk_yt}=e^{-2\pi i\vn_{k_xk_yt}\cdot\vec{\sigma}t/T},
	\label{eq:UN2}
\end{align}
where $\hat{n}_{k_xk_y t}$ is the unit vector along the $\vec{B}_{k_xk_y t}\times \hat{e}_z$ vector. The straightforward calculation gives
\begin{align}
	N_\text{Hopf}&=W_3[e^{-2\pi i\vn_{k_xk_yt}\cdot\vec{\sigma}t/T}]=1.
\end{align}
In other words, the adiabatic process~(\ref{eq:hN2}) is non-trivial $N=2$ Hopf
pump. Using the Bloch eigenvectors
\begin{align}
	\ket{u_{k_xk_ytn}}&=U_{k_xk_yt}\ket{n},
	\label{eq:ukN2}
\end{align}
with $n=1,2$, we find that the Berry connection $A_{\vk t}^n=A_{t}^n$ depends only on
time and the Chern-Simons 3-form is given by the area enclosed by the electron
\begin{equation}
	P_3^n=\frac{1}{2}\int_0^Tdt {\vec A}_{ t}^n
	\times\partial_t{\vec A}_{ t}^n =\frac{1}{2}. \label{eq:P3g}
\end{equation}
Therefore the two Chern-Simons 3-forms sum up to $1$, confirming the validity of the relation~(\ref{eq:6}).

\begin{figure}[t]
	\centering
	\includegraphics[width=\columnwidth]{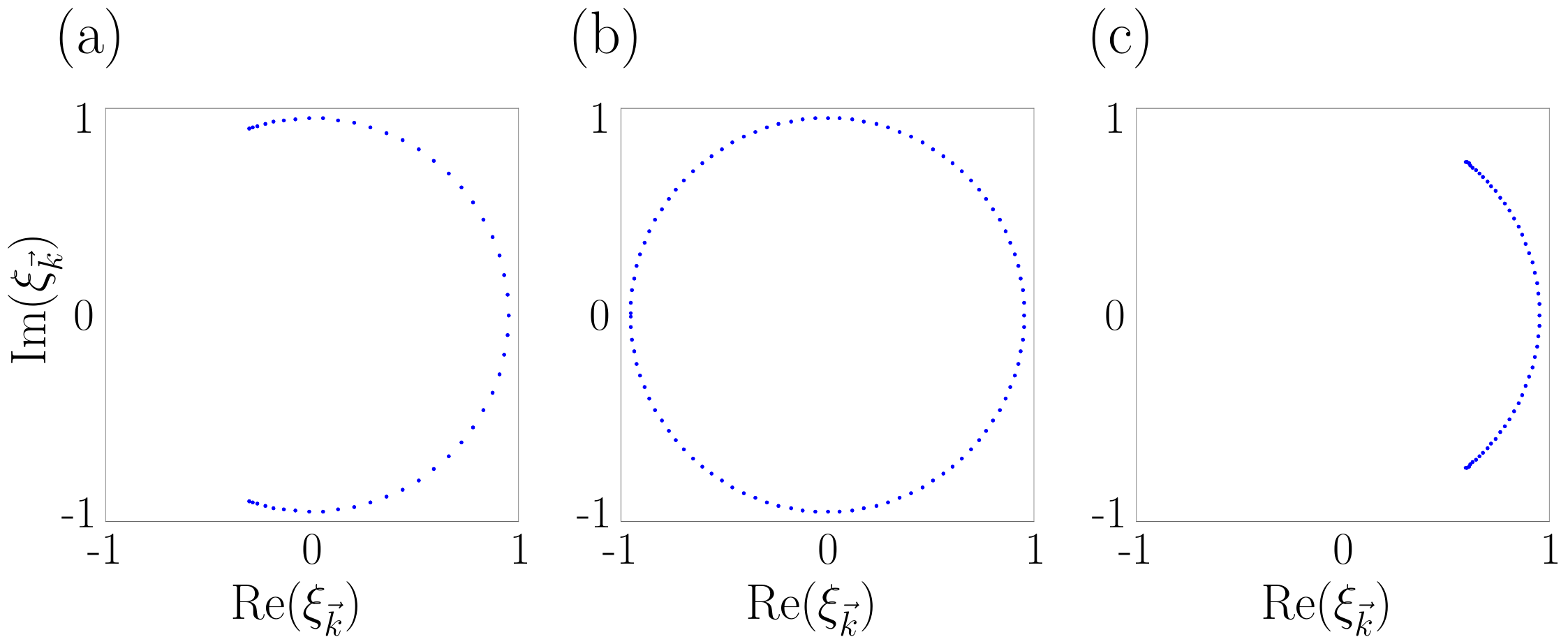}
	\caption{The spectrum $\xi_{\vec{k}}$ of the 1-cycle unitary operator $U^{\text{F}}_{k_xk_yT}$ defined in Eq.~\eqref{evolutionoperatornonadiab} for different values of $BT$. (a) $BT=3$. (b) $BT=5$, around which the quasienergy gap closes. (c) $BT=9$, after reopening of the quasienergy gap.}
	\label{fig:spectruuumbt}
\end{figure}

We now show that the time-dependent Hamiltonian~(\ref{eq:hN2}) is at the same
time an AFI, see Appendix~\ref{sec:floquet}. The time-evolution unitary
$U_{k_xk_yt}^\text{F}$ during each of the four segments is readily obtained as
\begin{align}
	U^\text{F}_{k_xk_yt}=&e^{-2\pi i\hat{n}_{k_xk_yt}\cdot\vec{\sigma} (t-t_0)/T}e^{-i(BT\sigma_3 -2\pi\hat{n}_{k_xk_yt}\cdot\vec{\sigma})(t-t_0)/T}.
\label{evolutionoperatornonadiab}
\end{align}
In the adiabatic limit, $BT\gg1$, the solution simplifies to
$U^\text{F}_{k_xk_yt}=e^{-2\pi i\hat{n}_{k_xk_yt}\cdot\vec{\sigma}
(t-t_0)/T}e^{-iB\sigma_3 (t-t_0)}$. As expected,~\cite{berry1984} the unitary
$U^\text{F}_{k_xk_yt}$ differs from the one in Eq.~(\ref{eq:UN2}) only by a dynamical
phase. Since in the adiabatic limit, $U^\text{F}_{k_xk_yT}=e^{-iBT\sigma_3}$, we
conclude that the model~(\ref{eq:hN2}) corresponds to Floquet insulator, which
remains to hold as long as $BT\gtrsim5$, see Fig.~\ref{fig:spectruuumbt}. Choosing $BT$ to be integer multiple
of $2\pi$, the relation $U^\text{F}_{k_xk_yT}=\sigma_0$ holds, and we find that
\begin{align}
	W_3[U^\text{F}_{k_xk_yt}]&=N_\text{Hopf}=1,
	\label{eq:W3UF}
\end{align}
i.e., the time-dependent Hamiltonian~(\ref{eq:hN2}) describes anomalous Floquet
insulator when $BT\gtrsim5$. See Appendix~\ref{sec:2banddetails} for details on the computation of $W_3[U^\text{F}_{k_xk_yt}]$.

\begin{figure}[t]
	\centering
	\includegraphics[width=\columnwidth]{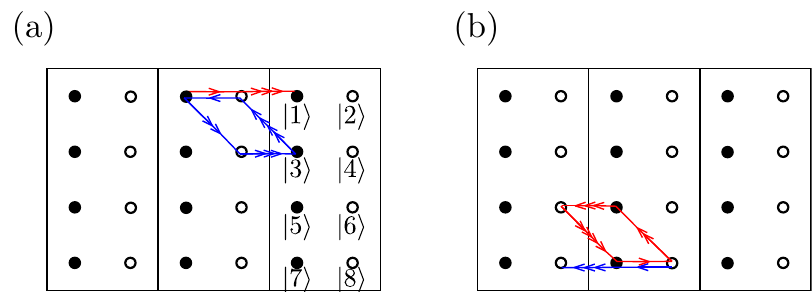}
	\caption{(a) Adiabatic time evolution of the upper edge subsystem
		$\mathcal{P}_{k_xt}^{\text{edge}}$. (b) The same as panel (a)
		for the lower edge subsystem
		$\mathcal{P}_{k_xt}^{\text{edge'}}$ (b). The orbital of the
		ribbon supercell are labeled by $\ket{n}$, $n=1,\dots,8$ as
	shown in panel (a).}
	\label{fig:pump}
\end{figure}

Lastly, we want to verify the bulk-boundary correspondence~\eqref{eq:bb}, by
computing explicitly the edge Chern number~\eqref{eq:C1Psurf}. We impose
open boundary conditions in the $y$-direction, and for concreteness take 4
layers in this direction, which defines the ribbon supercell of 8 orbitals
$\ket{n}$, $n=1,\dots,8$; see Fig.~\ref{fig:pump}. We note that we can consider such a narrow ribbon because in this model the WFs are highly localized. The WFs $\vert w_{R_xR_ytn}\rangle$ take the following form in the bulk
\begin{equation}
|w_{R_xR_yt1}\rangle=
\begin{cases}
\cos(t_0)|\vec{R} 1\rangle -\sin(t_0)|\vec{R} 2\rangle,\\
-\cos(t_{\frac{1}{4}})|\vec{R}2\rangle -\sin(t_{\frac{1}{4}})|\vec{R}+\vec{a}_y 1\rangle,\\
-\cos(t_{\frac{1}{2}})|\vec{R}+\vec{a}_y 1\rangle +\sin(t_{\frac{1}{2}})|\vec{R}+\vec{a}_y-\vec{a}_x 2\rangle,\\
\cos(t_{\frac{3}{4}})|\vec{R}\vec{a}_y-\vec{a}_x 2\rangle +\sin(t_{\frac{3}{4}})|\vec{R}1\rangle,
\end{cases}
\label{eq:WF1}
\end{equation}

\begin{equation}
|w_{R_xR_yt2}\rangle=
\begin{cases}
\cos(t_0)|\vec{R} 2\rangle +\sin(t_0)|\vec{R} 1\rangle,\\
\cos(t_{\frac{1}{4}})|\vec{R}1\rangle -\sin(t_{\frac{1}{4}})|\vec{R}-\vec{a}_y 2\rangle,\\
-\cos(t_{\frac{1}{2}})|\vec{R}-\vec{a}_y 2\rangle -\sin(t_{\frac{1}{2}})|\vec{R}-\vec{a}_y+\vec{a}_x 1\rangle,\\
-\cos(t_{\frac{3}{4}})|\vec{R}-\vec{a}_y+\vec{a}_x 1\rangle +\sin(t_{\frac{3}{4}})|\vec{R}2\rangle,
\end{cases}
\label{eq:WF2}
\end{equation}
where we used the notation $t_n=\frac{2\pi}{T}(t-nT)$ and the four expressions
for the bulk WFs correspond to the four time-segments as in Eq.~(\ref{eq:S4}). The
Fourier transform
along the $x$-direction gives the hybrid bulk WFs $\vert w_{k_xR_ytn}\rangle$,
that are used to compute the edge projector
$\mathcal{P}_{k_xt}^{\text{edge}}$ given in Eq.~\eqref{eq:Psurf}. We perform a Wannier cut
by removing four bulk WFs from the middle of the ribbon, followed by projecting
onto the upper half of the ribbon supercell. The only nonzero contributions to
the edge Chern number come from the time-segments $t\in [\frac{T}{2},T)$. For
$t\in[\frac{T}{2},\frac{3T}{4})$, $\mathcal{P}_{k_xt}^{\text{edge}}=\text{diag}(A,A)$, with
\begin{equation}
A=\begin{pmatrix}
	\sin(t_\frac{1}{2})^2 & e^{-ik_x}\cos(t_\frac{1}{2})\sin(t_\frac{1}{2})\\
	e^{ik_x}\cos(t_\frac{1}{2})\sin(t_\frac{1}{2})& \cos(t_\frac{1}{2})^2\\
\end{pmatrix},
\end{equation}
where the basis $\{\ket{1},\ket{2},\ket{3},\ket{4}\}$ has been used to write
${\cal P}_{k_xt}^{\text{edge}}$. In this case, we obtain that $\text{Tr}\left(
{\cal P}^{\text{edge}}_{k_xt}[\partial_{k_x}{\cal
P}^{\text{edge}}_{k_xt},\partial_{t} {\cal P}^{\text{edge}}_{k_xt}]_-\right)=
\frac{4\pi i}{T}\sin\left(\frac{4\pi t}{T}\right)$. Similarly, for $t\in
[\frac{3T}{4},T)$,
\begin{multline}
\mathcal{P}_{k_xt}^{\text{edge}}=\\\begin{pmatrix}
1 & 0 & 0 & 0\\
0 &	\sin(t_{\frac{3}{4}})^2 & -e^{ik_x}\cos(t_\frac{3}{4})\sin(\frac{3}{4}) & 0 \\
0 & -e^{-ik_x}\cos(t_\frac{3}{4})\sin(t_\frac{3}{4}) & \cos(t_\frac{3}{4})^2 & 0 \\
0 & 0 & 0 & 0
\end{pmatrix},
\end{multline}
we obtain that $\text{Tr}\left( {\cal
P}^{\text{edge}}_{k_xt}[\partial_{k_x}{\cal
P}^{\text{edge}}_{k_xt},\partial_{t} {\cal P}^{\text{edge}}_{k_xt}]_-\right)=
\frac{2\pi i}{T}\sin\left(\frac{4\pi t}{T}\right)$. Therefore
$\text{Ch}^{\text{edge}}=-1$, confirming the bulk-boundary
correspondence~(\ref{eq:bb}). 

\subsubsection{$N=3$ band Hopf pump}\label{sec:N3Hopfp}
In this example, we consider a $N=3$ band model that is obtained from the $N=2$ band
model, introduced in the previous subsection, after adding an additional
orbital in the unit cell. Furthermore, we introduce a parameter $\delta$ in the
model, such that $\delta=0$ corresponds to the previously discussed $N=2$ Hopf
insulator with the additional orbital not being involved in the adiabatic
process. This way, for $\delta=0$ we have
$P_3^1=P_3^2=\frac{1}{2}$, while $P_3^3=0$. For $\delta\neq 0$ the model is
chosen to have the property $P_3^1\neq P_3^2$ (and $P_3^3=0$), as we discuss below.

\begin{figure}[t]
	\centering
	\includegraphics[width=.8\columnwidth]{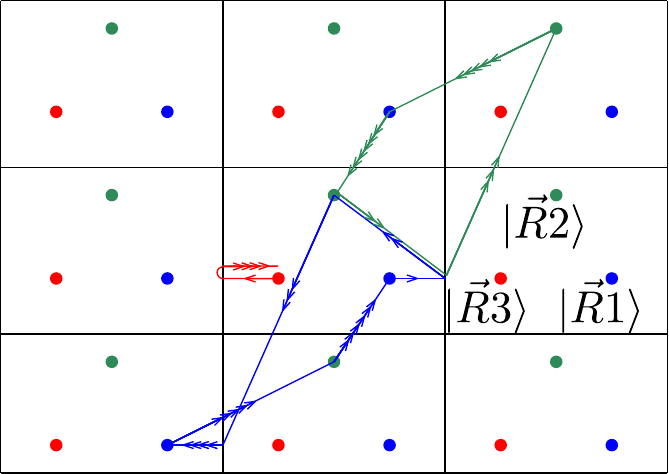}
	\caption{Three level periodic drive made of 6 steps of equal duration.
		At the first and fourth stages, we consider an incomplete
		rotation between the orbitals $\ket{\vR1}$ and
		$\ket{(\vR+\va_x)3}$, such that the area enclosed by the trajectory of the third orbital is zero. This is controlled by the parameter
	$\delta\in[0,1)$.} \label{fig:4}
\end{figure}

We consider a driven model with three sites per unit cell, labeled by
$\{\ket{\vR 1},\ket{\vR 2},\ket{\vR3}\}$. The driving protocol has the period
$T$ and is made of 6 time segments of equal duration $\frac{T}{6}$, which are
illustrated in Fig.~\ref{fig:4}.

The eigenvalues of the Hamiltonian $h_{k_xk_yt}$ are chosen to be
time-independent and take the values $1,2,3$
\begin{align}
	h_{k_xk_yt}&=\sum_{n=1}^3 n\ket{u_{k_xk_ytn}}\bra{u_{k_xk_ytn}},
	\label{eq:hN3}
\end{align}
where the Bloch eigenvectors $\ket{u_{k_xk_ytn}}$ can be read off from Fig.~\ref{fig:4} and are explicitly given in Appendix~\ref{sec:3banddetails}.
At each of the six time segments,
the adiabatic process involves only two orbitals. For example, during the first segment, see
Fig.~\ref{fig:4}, the orbitals involved are $\ket{\vR1}$ and
$\ket{(\vR+\va_x)3}$, where the final state is a $\delta$-dependent superposition
of these two states. Explicit calculation gives
\begin{equation}
\begin{cases}
P_3^{1}=\frac{1}{2}\left(1-\sin^2\left(\frac{\pi}{2}\delta\right)\right)\leq\frac{1}{2},\\
P_3^{2}=\frac{1}{2}\left(1+\sin^2\left(\frac{\pi}{2}\delta\right)\right)\geq\frac{1}{2},\\
P_3^{3}=0.
\end{cases}
\end{equation}
The contributions $P_3^{1}$ and $P_3^{2}$ are not anymore quantized and equal
 if $\delta\neq0$. However the sum turns out to be quantized and equal
to $1$, just like in the previous example.  The winding number can be
computed along the same lines as in the previous section, and is given by $N_\text{Hopf}=W\left[
U_{\vk t} \right]=1$. Since $N_\text{Hopf}=\sum_{n=1}^N P_3^n$ holds, we conclude that the non-topological orbital magnetization~\cite{trifunovic2019a} $\sum_{n=1}^N m_n^\text{nontop}$ vanishes for this example.

\section{Conclusions}\label{sec:conclusions}
In this work we explore ``beyond tenfold-way'' topological phases that
belong to the category of delicate multi-gap phases. These phases can be band insulators
with $(N-1)$ band gaps, where the number of bands between two successive band
gaps is fixed. We obtain a $\ZZ$ classification for
three-dimensional band insulators without any symmetry constraints. Unlike the tenfold-way topological insulators, the $N$-band Hopf insulator does not host topologically protected gapless modes. Since both the bulk and the boundary are fully gapped, and there is no unique way to separate a finite system into bulk and boundary subsystems.~\cite{trifunovic2020,ren2021} Despite this non-uniqueness, we formulate the bulk-boundary correspondence stating that a finite sample of the $N$-band Hopf insulator
consists of the bulk, with the total magnetoelectric polarizability of all the bulk bands quantized to an integer value, wrapped in a sheet of a Chern insulator with the total Chern number of all the boundary bands equal to minus the same integer value (see Fig.~\ref{fig:ball}). The obtained classification and bulk-boundary correspondence is the same as that of the Hopf insulator (the case $N=2$). Hence, we dub these new phases the $N$-band Hopf insulators.

An ultimate ``usefulness test'' for the beyond-tenfold-way
classifications is whether the obtained topological phases are accompanied by a
quantized boundary effect. While in the tenfold-way classification the chemical
potential $\mu$ plays a crucial role, since only the bands with the energy below
$\mu$ are classified, and the quantized boundary effect can be observed in
equilibrium, the quantized boundary effect in the multi-gap topological phases can only
be observed  out of equilibrium. This is because the multi-gap
classifications schemes, both the stable and the delicate ones, classify the whole
band structure and the chemical potential plays no role. Hence, one would
naturally attempt to fill all the bands with electrons in order to
observe a quantized effect. We find that if the whole finite sample is fully
filled with electrons, the quantized bulk and boundary effects mutually cancel.
Therefore, the quantized boundary effect of the $N$-band Hopf insulator can be
observed in a non-equilibrium state where only the states close to the
boundary are fully filled with electrons.

Our work discusses not only the three-dimensional $N$-band Hopf insulators but also the two-dimensional $N$-band Hopf pumps. The Hopf pump can be seen as a bulk with the quantized (geometric) orbital magnetization~\cite{trifunovic2019a}, with a Thouless pump (of all the bands combined) at the edges. Furthermore, we discuss a particular example of the Hopf pump that illustrates its similarities with the anomalous Floquet insulator.~\cite{rudner2013} On the other hand, the difference between the $N$-band Hopf pump, introduced in this work, and the anomalous Floquet insulator is that the latter requires the gap in the quasienergy spectrum which is difficult to
guarantee experimentally, whereas having a multi-gap band structure is more physical requirement. Furthermore, the stable multi-gap classification (i.e., Floquet insulators) always results in Abelian groups, while it has been
shown~\cite{wu2019} that the delicate multi-gap classification of one-dimensional band insulators with certain magnetic point-group symmetry is non-Abelian. Thus, extending the program outlined in this work to systems of different dimensions and with additional symmetry constraints is highly desirable.

\subsection*{Acknowledgement}
The authors thank Aris Alexandradinata, Tom{\' a}{\v s} Bzdu{\v s}ek, Alexandra Nelson, David Vanderbilt, and Haruki Watanabe for fruitful discussions. BL acknowledges funding from the European Research Council (ERC) under the European Union’s Horizon 2020 research and innovation program (ERC-StG-Neupert-757867-PARATOP). LT acknowledges financial support from the FNS/SNF Ambizione Grant No.~PZ00P2\_179962.

\appendix
\section{Stable multi-gap classification of two-dimensional Floquet insulators}\label{sec:floquet}
Periodically driven band structure $h^\text{F}_{k_xk_yt}$ is example of
Floquet system. The Floquet system is called insulator, if the spectrum
(quasienergy spectrum $e^{i\varepsilon_{k_xk_y}T}$) of the unitary Floquet
operator $U^\text{F}_{k_xk_yT}={\cal T}e^{-i\int_0^T
h^\text{F}_{k_xk_yt}dt}$ has at least one gap on the unit circle in the
complex plane. Below we review classification of two-dimensional Floquet
insulators.

Topological classification of Floquet insulators, classifies the unitary
evolution operator $U^\text{F}_{k_xk_yt}$ under the constraint that one or
multiple gaps in the quasienergy spectrum are maintained. In other words, two
Floquet insulators are said to be topologically equivalent if their unitary
evolution operator can be brought to the same form without closing the gap
(gaps) in the quasienergy spectrum. Mathematically, such constraint divides the
total Hilbert space into mutually orthogonal subspaces with corresponding
projectors ${\cal P}^\text{F}_{k_xk_yn}$ spanned by the eigenvectors of the
Floquet operator $U^\text{F}_{k_xk_yT}$ with quasienergies between the two
neighbouring gaps.  One typically considers $K$-theoretic (i.e. stable)
classification where the ranks of the projectors ${\cal P}^\text{F}_{k_xk_yn}$
can be varied by addition of trivial quasibands. For $N$ gaps in quasienergy
spectrum, two-dimensional Floquet insulators have $\ZZ^N$
classification.~\cite{roy2017} The subgroup $\ZZ_\text{Chern}^{N-1}\subset\ZZ$
is generated by $N-1$ Chern numbers corresponding to subspaces ${\cal
P}^\text{F}_{k_xk_yn}$, for say $n=1,\dots,N-1$. The remaining $\ZZ$
topological invariant is given by the third winding number
$W_3[U_{k_xk_yt}^{\text{F},\varepsilon}]$ of the unitary
$U_{k_xk_yt}^{\text{F},\varepsilon}$ which is obtained from the unitary evolution
$U_{k_xk_yt}^\text{F}$ by continuously deforming the Floquet operator
$U_{k_xk_yT}^\text{F}$ to identity matrix while maintaining the gap around some
quasienergy $\varepsilon$ ($\varepsilon$ belongs to one of the $N$ 
gaps in quasienergy spectrum).

The Floquet insulators with topological invariants from the subgroup
$\ZZ_\text{Chern}^{N-1}$ can all be realized as static systems. The remaining
generator which is diagnozed by the third winding number exists only for
time-dependent band structures and is called anomalous Floquet
insulator.~\cite{rudner2013,titum2016}

Anomalous Floquet insulator was found to obey bulk-boundary
correspondence.~\cite{titum2016} Consider anomalous Floquet insulator that
satisfies $U_{k_xk_yT}=\mathbbm{1}_{N\times N}$. We apply open boundary
conditions in $y$-direction and consider slab geometry with $N_y$ layers, with
time-dependent band structure $h^\text{F}_{k_xt}$ and $NN_y\times NN_y$
unitary evolution operator $U_{k_xt}^{\text{F},\text{slab}}$. The
bulk-boundary correspondence states
\begin{align}
\label{eq:15}
	N_\text{AFI}&\equiv
	W_3[U^\text{F}_{k_xk_yt}]\\
	&=\int_0^{2\pi}\frac{dk_x}{2\pi}\text{Tr}[U_{k_xt}^{\text{F},\text{slab}\dagger}\partial_{k_x}U_{k_xt}^{\text{F},\text{slab}}\theta(y)\theta(y^\prime)].\nonumber
\end{align}

In other words, anomalous Floquet insulator with $N_\text{AFI}\neq0$ induces
quantized charge pumping of $N_\text{AFI}$ electrons along the
boundary in steady state.~\cite{titum2016,kundu2020}

\section{Derivation of the relation~\eqref{eq:6}}\label{sec:eq6}
In this appendix we derive relation~\eqref{eq:6}, following closely the derivation presented in Ref.~\onlinecite{unal2019}. We denote the set of orbitals in the unit cell by $\{\ket{1},\ket{2}\}$, and the unitary matrix transforming this basis to the Bloch eigenvectors $\{\ket{u_{\vk 1}},\ket{u_{\vk,2}}\}$ by $U_{\vk}$, i.e., $\ket{u_{\vk a}}=U_{\vk}\ket{a}$, with $a=1,2$. The winding number~\eqref{eq:5} of the unitary matrix $U_{\vk}$ can be written in the following form
\begin{equation}
W_3[U_{\vk}] =\int_{\text{BZ}} \frac{d^3k}{24\pi^2} \epsilon^{ijk}\sum_{a,b,c,d}U^{\dagger}_{ab}\partial_i U_{bc} \partial_j U^{\dagger}_{cd} \partial_k U_{da},
\end{equation}
where indices $i$ and $j$ run over $k_x,k_y,k_z$, whereas $a,b,c,d$ run over the two band indices, and the summation over repeated indices is assumed.. The matrix elements of the unitary $U_{\vk}$ are defined as $U_{ab}(\vk)=\bra{a}U_{\vk}\ket{b}$.
\begin{multline}
W_3[U_{\vk}] =\\ \int_{\text{BZ}} \frac{d^3k}{24\pi^2} \epsilon^{ijk}\sum_{a,b,c,d}\langle u_{\vk a}\vert b\rangle\partial_i \langle b\vert u_{\vk c}\rangle \partial_j \langle u_{\vk c}\vert d\rangle \partial_k \langle d\vert u_{\vk a}\rangle.
\end{multline}
We note that the derivatives act only on the Bloch eigenstates, which leads to
\begin{equation}
W_3[U_{\vk}] =\int_{\text{BZ}} \frac{d^3k}{24\pi^2} \epsilon^{ijk}\sum_{a,c}\langle u_{\vk a}\vert \partial_i u_{\vk c}\rangle \langle \partial_j u_{\vk c} \vert\partial_k u_{\vk a}\rangle.
\end{equation}
Taking the summation over the two bands, we obtain four terms 
\begin{equation}
W_3[U_{\vk}] = \int_{\text{BZ}} \frac{d^3k}{24\pi^2} \epsilon^{ijk}[W_1+W_2+W_3+W_4],
\end{equation}
where
\begin{equation}
\begin{cases}
W_1=u_{\vk 1}^{\dagger}\partial_i u_{\vk 1}\partial_j u_{\vk 1}^{\dagger} \partial_k u_{\vk 1},\\
W_2=u_{\vk 1}^{\dagger}\partial_i u_{\vk 2}\partial_j u_{\vk 2}^{\dagger} \partial_k u_{\vk 1},\\
W_3=u_{\vk 2}^{\dagger}\partial_i u_{\vk 1}\partial_j u_{\vk 1}^{\dagger} \partial_k u_{\vk 2},\\
W_4=u_{\vk 2}^{\dagger}\partial_i u_{\vk 2}\partial_j u_{\vk 2}^{\dagger} \partial_k u_{\vk 2}.\\
\end{cases}
\end{equation}
By following the procedure outlined in Ref.~\onlinecite{unal2019}, the first and fourth terms give an equal contribution, $W_1=W_4$, while the second and third term give also an equal contribution, $W_2 = W_3 = 2W_1$. Therefore we conclude that 
\begin{equation}
W_3[U_{\vk}] =2\int_{\text{BZ}}\frac{d^3k}{8\pi^2}\epsilon^{ijk}u_{\vk 1}^{\dagger}\partial_i u_{\vk 1}\partial_j u_{\vk 1}^{\dagger} \partial_k u_{\vk 1}
\end{equation}
using the definition of the Berry connection, $(\vA_n)_j=i\langle u_{\vk n}\vert\partial_{j} u_{\vk n}\rangle$, we note that $\epsilon^{ijk}(\vA_n)_i\partial_j(\vA_n)_k = \vA_n\cdot\vec\nabla\times\vA_n$. We finally recover the expression for Abelian third Chern-Simons form, hence we conclude that $W_3[U_{\vk}]=P_3^1+P_3^2=2P_3^1$ holds.

\section{Delicate multi-gap classification of one-dimensional real band structures}\label{sec:nonstable1d}
We consider the delicate multi-gap topological classification of one-dimensional
real Hamiltonians using the method of Sec.~\ref{sec:Nhopf} of the main text.
The resulting classification group is non-Abelian as first discussed by Wu,
Soluyanov, and Bzdu{\v s}ek.~\cite{wu2019} Here we show that the classification method used in the main text also gives the expressions for strong topological invariants that were previously not known.

As we show below, the delicate multi-gap classification of real one-dimensional band structures depends explicitly on the number of bands $N$ (i.e. gaps). Hence, unlike the case of $N$-band Hopf insulators discussed in the main text, the group structure is given by concatenation of two Bloch Hamiltonians $h^{(1)}_k$ and $h^{(2)}_k$ with the same number of bands
\begin{align}
	h^{(2)}_k\circ h^{(1)}_k=
	\begin{cases}
		h^{(1)}_{2k} & \text{for } k\in[0,\pi),\\
		h^{(2)}_{2k-2\pi} & \text{for } k\in[\pi,2\pi).
	\end{cases}
	\label{eq:concat}
\end{align}
In order for the
concatenated Hamiltonian to be continuous, we require that $h^{(1)}_0=h^{(2)}_0$,
which can be always achieved by deformation given that there are no weak
topological invariants.

The flattened Bloch Hamiltonian is diagonalized
\begin{align}
	h_k=O_k \text{diag}(1,\dots,N)O_k^T,
	\label{eq:hkreal}
\end{align}
where $O_k\in SO(N)$ is assumed continuous for $k\in[0,2\pi)$. Note that the
periodicity of the Bloch Hamiltonian $h_0=h_{2\pi}$, does not require
$O_0=O_{2\pi}$, but rather weaker requirement $O_0^TO_{2\pi}\in
O(1)^{N-1}\subset SO(N)$. In other words, the real Bloch eigenvectors
$\ket{u_{kn}}$ do not need to be continuous at $k=2\pi$ but
$\ket{u_{0n}}=\pm\ket{u_{2\pi n}}$.  We now define auxiliary orthogonal matrix
$o(k)$
\begin{align}
	o_k&=O_0^TO_{k}.
	\label{eq:ok}
\end{align}
The strong classification of the real Hamiltonians~(\ref{eq:hkreal}) is obtained by
classifying orthogonal matrices $o_k$, since the orthogonal matrix $O(0)$
contains only weak invariants. We have that $o_0=\mathds{1}_{N\times N}$ and
$o_{2\pi}\in O(1)^{N-1}$, i.e., $o_k$ is classified by the relative homotopy
group $\pi_1(SO(N),O(1)^{N-1})$. The group $\pi_1(SO(N),O(1)^{N-1})$ can be
found using the following exact sequence
\begin{align}
	\pi_1(O(1)^{N-1})&\xrightarrow{i_1}\pi_1(SO(N))\xrightarrow{i}\pi_1(SO(N),O(1)^{N-1})\nonumber\\
	&\xrightarrow{\partial}\pi_0(O(1)^{N-1})\xrightarrow{i_0}\pi_0(SO(N)),
	\label{eq:exactsec1}
\end{align}
The groups $\pi_1(O(1)^{N-1})$ and $\pi_0(SO(N))$ are trivial. We first
consider case $N>2$ (for the $N=2$ case see Sec.~\ref{sec:N2}), where
$\pi_1(SO(N))=\ZZ_2$ and $\pi_0(O(1)^{N-1})=\ZZ_2^{N-1}$ holds,
\begin{align}
	0\xrightarrow{i_1}\ZZ_2\xrightarrow{i}\pi_1(SO(N),O(1)^{N-1})\xrightarrow{\partial}\ZZ_2^{N-1}\xrightarrow{i_0}0.
\end{align}
The above extension problem does not have an unique solution, i.e., there is
more than one group that satisfies the above exact sequence if the
homomorphisms $i$ and $\partial$ are not specified. We show in
Sec.~\ref{sec:gstruct} that $\pi_1(SO(N),O(1)^{N-1})$ is non-Abelian group.

To each element of the group $\pi_1(SO(N),O(1)^{N-1})$, that is represented by
some path $o_k$ defined by relations~(\ref{eq:ok}) and~(\ref{eq:hkreal}), we
can assign topological invariants from the Abelian groups $\pi_1(SO(N))=\ZZ_2$ and
$\pi_0(O(1)^{N-1})=\ZZ_2^{N-1}$.  The $\ZZ_2$ topological invariants $\nu_i$ for
$i=1,\dots,N-1$ are defined as follows
\begin{align}
	\nu_i&=\mathrm{sign}[ (o_{2\pi})_{ii}].
	\label{eq:nui}
\end{align}
To assign a $\ZZ_2$ topological invariant $\mathfrak{p}$ to an arbitrary path in
$\pi_1(SO(N),O(1)^{N-1})$, we need a convention that assigns a loop to the given
path, because $\pi_1(SO(N))$ is defined for loops only. Using the vector
notation $\bm \nu$, where $(\bm\nu)_m=\nu_m$ for $i=1,\dots,N-1$, we define
reference paths $o^{\bm e_m}_{\text{ref},k}$ for $m=1,\dots,N-1$
\begin{align}
	o^{\bm e_m}_{\text{ref},k}=e^{i\hat L_{mN} k/2},
	\label{eq:oref}
\end{align}
where $(\hat L_{ij})_{mn}=-i(\delta_{im}\delta_{nj}-\delta_{jm}\delta_{in})$ are
generators of $SO(N)$. $o^{\bm e_m}_{\text{ref},2\pi}$ is the $\pi$ rotation in
the plane spanned by Bloch eigenvectors $\ket{u_{0m}}$ and $\ket{u_{0N}}$.  For an arbitrary value
of the topological invariants from the righthand side of the exact
sequence~(\ref{eq:exactsec1}), $\bm \nu=\bm e_m+\bm e_{m^\prime}+\dots$, we
define
\begin{align}
	o^{\bm \nu}_{\text{ref},k}=o^{\bm e_m}_{\text{ref},k}\circ o^{\bm e_{m^\prime}}_{\text{ref},k}\circ\dots,
	\label{eq:orefg}
\end{align}
where the concatenation is ordered from the smallest to the largest index
$m>m^\prime>\dots$. Hence, a loop of orthogonal matrices $o^L_k$ can be
uniquely assigned to each $o_k$, that has the topological invariants $\bm \nu$,
\begin{align}
	o^L_k=o_k\circ(o^{\bm \nu}_{\text{ref},k})^{-1},
	\label{eq:oL}
\end{align}
where the notation $(o_k)^{-1}=o_{2\pi-k}$ has been used. To each $o^L_k$ an
element $\mathfrak{p}\in\{-1,1\}$ from $\pi_1(SO(N))$ can be assigned, as we
review in Sec.~\ref{sec:pi1SO}. Therefore, each element of
$\pi_1(SO(N),O(1)^{N-1})$ can be specified by topological invariants
$\mathfrak{p}$ and $\bm\nu$. In Sec.~\ref{sec:gstruct} using the concatenation
of the matrices $o_k$ we show that the groups structure of
$\pi_1(SO(N),O(1)^{N-1})$ is non-Abelian.

\subsection{The topological invariant of $\pi_1(SO(N))$}\label{sec:pi1SO}
To each loop $o^L_k\in SO(N)$, $o^L_0=o^L_{2\pi}=\mathds{1}$, we need to
assign (continuously) an element $\bar o_k\in Spin(N)$.  After such
assignment, the $\ZZ_2$ topological invariant $\mathfrak{p}$ is given by
\begin{align}
	\bar o^L_{2\pi}=\mathfrak{p}\mathds{1}.
	\label{eq:p}
\end{align}
To obtain $\bar o^L_k$, we need $N$ Dirac matrices $\gamma_m$ for
$m=1,\dots,N$ (the dimension of the representation is unimportant). The Dirac
matrices satisfy the following algebra
\begin{align}
	\gamma_m^2&=1,\nonumber\\
	\gamma_m\gamma_n&=-\gamma_n\gamma_m,\text{ for }m\neq n.
	\label{eq:dirac}
\end{align}
Consider a grid with $M$ points in the Brillioun zone, where each segment
of the grid has length $\delta k$, $M=2\pi/\delta k$. The rotation loop
$o^L_k$ can be approximated by series of rotations around the
piecewise fixed axes, i.e.,
\begin{align}
	o^{L}_{m\delta k}&=o^{L,(m)}_{\delta k}o^{L,(m-1)}_{\delta k}\dots o^{L,(1)}_{\delta k},
	\label{eq:grid}
\end{align}
for $m=1,\dots,M$, where each orthogonal matrix $o^{L,(n)}_{\delta k}$ represents rotation around the
fixed axis
\begin{align}
	o^{L,(n)}_{\delta k}=e^{i\sum \hat L_{ab} \theta_{ab}}.
\end{align}
The axes (labeled by the indices $a$ and $b$) and the angles $\theta_{ab}$ are
found by diagonalizing $o^{L,(n)}_{\delta k}$. The matrix $\bar o^L_{2\pi}$ is found
by replacing each $o^{L,(n)}_{\delta k}$ in the product~(\ref{eq:grid}) by $\bar
o^{L,(n)}_{\delta k}$
\begin{align}
	\bar o^{L,(n)}_{\delta k}=e^{i\sum [\gamma_a,\gamma_b]  \theta_{ab}/4}.
	\label{eq:spinn}
\end{align}

For practical purposes, one can more easily compute $\mathfrak{p}$ as follows.~\cite{trifunovic2017}
The eigenvalues of $o^L_k$ come in pairs $e^{\pm i\theta_{kn}}$ (otherwise
they are real). By plotting the phases $\pm\theta_{kn}$ between $[-\pi,\pi]$, we
can find $\mathfrak{p}$ by counting the number of crossings (on real axis)
modulo two. 

\subsection{The group structure of $\pi_1(SO(N),O(1)^{N-1})$}\label{sec:gstruct}
For each $o_k$ defined by relations~(\ref{eq:ok}) and~(\ref{eq:hkreal}), we
can compute $N$ topological invariants $\bm \nu$ and $\mathfrak{p}$, as
discussed above. We use the convention that $\nu_m\in\{0,1\}$ whereas
$\mathfrak{p}\in\{-1,1\}$. We map an element of $\pi_1(SO(N),O(1)^{N-1})$ to
the following string of Dirac matrices
\begin{align}
	\mathfrak{p}(i\gamma_1)^{\nu_1}(i\gamma_2)^{\nu_2}\dots(i\gamma_{N-1})^{\nu_{N-1}}.
	\label{eq:mapping}
\end{align}
Below we prove that above map is an isomorphism. To this end, we need to show
that the concatenation satisfies the algebra~(\ref{eq:dirac}). Consider first
$o^{\bm e_m}_{\text{ref},k}$ which is mapped to $i\gamma_m$ (by the
construction~(\ref{eq:oref}) it has $\mathfrak{p}=1$ and $\bm \nu=\bm e_m$).
The element $o^{\bm e_m}_{\text{ref},k}\circ o^{\bm e_m}_{\text{ref},k}$ is a loop
(i.e. it is equal to the identity matrix for $k=2\pi$), thus $\bm \nu=0$.
Additionally, we have $\mathfrak{p}=-1$ since $o^{\bm e_m}_{\text{ref},k}\circ
o^{\bm e_m}_{\text{ref},k}$ represents rotation by $2\pi$ angle in the plane
spanned by the Bloch eigenvectors $\ket{u_{0m}}$ and $\ket{u_{0N}}$, see Eq.~(\ref{eq:oref}). We conclude that
$o^{\bm e_m}_{\text{ref},k}\circ
o^{\bm e_m}_{\text{ref},k}$ should be mapped to $-\mathds{1}$ which is in
agreement with $(i\gamma_m)^2=-\mathds{1}$. Next consider an element $o^{\bm
e_m}_{\text{ref},k}\circ o^{\bm e_n}_{\text{ref},k}$ with $m>n$, which has
topological invariants $\bm \nu=\bm e_n+\bm e_m$ and $\mathfrak{p}=1$ and is
mapped to $(i\gamma_n)(i\gamma_m)$. On the other hand, an element $o^{\bm
e_n}_{\text{ref},k}\circ o^{\bm e_m}_{\text{ref},k}$ has the same topological
invariants $\bm \nu=\bm e_n+\bm e_m$, and rule~(\ref{eq:oL}) assigns the
following loop to it
\begin{align}
	o^L_k=o_{\text{ref},k}^{\bm e_n}\circ o_{\text{ref},k}^{\bm e_m}\circ (o_{\text{ref},k}^{\bm e_n})^{-1}\circ (o_{\text{ref},k}^{\bm e_m})^{-1}.
\end{align}
The mapping~(\ref{eq:spinn}) sends $o^L_{2\pi}$ to
\begin{align}
	\bar o^L_k=e^{i\sigma_1\pi}e^{i\sigma_2\pi}e^{-i\sigma_1\pi}e^{-i\sigma_2\pi}=-\sigma_0,
\end{align}
which implies $\mathfrak{p}=-1$, see Eq.~(\ref{eq:p}). Thus, $o^{\bm
e_n}_{\text{ref},k}\circ o^{\bm e_m}_{\text{ref},k}$ is mapped to
$-(i\gamma_n)(i\gamma_m)=(i\gamma_m)(i\gamma_n)$, which proves that the
considered map is an isomorphism between $\pi_1(SO(N),O(1)^{N-1})$ and
the algebra~(\ref{eq:dirac}).

\subsection{The two-band case $N=2$}\label{sec:N2}
This case is special because $\pi_1(SO(2))=\ZZ$, thus the exact sequence~(\ref{eq:exactsec1}) reads
\begin{align}
	0\rightarrow\ZZ\rightarrow\pi_1(SO(2),O(1))\rightarrow\ZZ_2\rightarrow0.
\end{align}
In the same way as previously, we can assign two topological invariants to
$o_k\in SO(2)$ that we denote by $\mathfrak{N}\in\ZZ$ and $\nu$. To define
$\mathfrak{N}$ we use $o_{\text{ref},k}$ (in $N=2$ case there is only one
reference rotation) to define the corresponding loop $o^L_k$, to which we can associate the
winding number. It is easy to check that in this case, $o_{\text{ref},k}$ with
the topological invariants $\mathfrak{N}=0$ and $\nu=1$ generates the whole group
$\pi_1(SO(2),O(1))$: the concatenation of $o_{\text{ref},k}$ $n$-times
with itself gives an element with the topological invariants $\mathfrak{N}=\lfloor
n/2 \rfloor$ and $\nu=n\mod 2$. Thus $\pi_1(SO(2),O(1))=\ZZ$ where
the topological invariant is equal to $\theta_{2\pi}/\pi$, with
$\theta_k\in\mathbb{R}$ being the angle of rotation~\footnote{The group
$\mathbb{R}$ of real numbers is universal cover of the group
$\pi_1(SO(2))=S^1$.} associated to $o_k\in SO(2)$.

\section{Two-dimensional $N$-band Hopf pump}
Below we give details of the calculations for $N=2$ and $N=3$ Hopf pump.

\subsection{The $N=2$ Hopf pump}\label{sec:2banddetails}
We compute the winding number for the 2-band model in the adiabatic limit $BT\gg 1$. In this limit, the evolution operator takes the form $U^\text{F}_{k_xk_yt}=e^{-2\pi i\hat{n}_{k_xk_yt}\cdot\vec{\sigma}
(t-t_0)/T}e^{-iB\sigma_3 (t-t_0)}$. This operator takes the form
\begin{equation}
U^\text{F}_{k_xk_yt}=\begin{cases}e^{-2\pi i\sigma_2
t/T}e^{-iB\sigma_3 t}\\
-ie^{-2\pi i(\sin(k_y)\sigma_1+\cos(k_y)\sigma_2)t_{\frac{1}{4}}}\\\quad\times e^{-iBT\sigma_3 t_{\frac{1}{4}}}\sigma_2e^{-i\frac{BT}{4}\sigma_3}  \\
-e^{-2\pi i(\sin(k_x)\sigma_1+\cos(k_x)\sigma_2)t_{\frac{1}{2}}}\\\quad\times e^{-iBT\sigma_3 t_{\frac{1}{2}}}e^{ik_y\sigma_3}
\\
e^{-2\pi i(\sin(\delta k)\sigma_1+\cos(\delta k)\sigma_2)t_{\frac{3}{4}}}\\\quad\times e^{-iBT\sigma_3 t_{\frac{3}{4}}}\sigma_2e^{-i(\delta k+\frac{BT}{4}) \sigma_3}
\end{cases}
\end{equation}
where we introduced the notation $\delta k =k_x-k_y$.
We notice that only during the third segment of the drive the unitary $U^\text{F}_{k_xk_yt}$ will give a non trivial contribution to the winding number, as it does not depend independently on $k_x$ and $k_y$ for the other segments of the drive. One then finds
\begin{equation}
(U^\text{F}_{k_xk_yt})^{\dagger}\partial_{k_y}U^\text{F}_{k_xk_yt}=i\sigma_3
\end{equation}

\begin{multline}
(U^\text{F}_{k_xk_yt})^{\dagger}\partial_{k_x}U^\text{F}_{k_xk_yt}=-i\sin^2\left(\frac{2\pi t}{T}\right)\sigma_3
\\ -\frac{i}{2}\sin\left(\frac{4\pi t}{T}\right)[\cos(a)\sigma_1+\sin(a)\sigma_2]
\end{multline}
\begin{equation}
(U^\text{F}_{k_xk_yt})^{\dagger}\partial_{t}U^\text{F}_{k_xk_yt}=-i\frac{BT}{T}\sigma_3 -\frac{2\pi}{T} [\cos(a)\sigma_1-\sin(a)\sigma_2]
\end{equation}
where we introduced the notation $a=k_x - 2 k_y + (-1+\frac{2t}{T})BT$.
We conclude that the trace gives $\frac{4\pi}{T}\sin\left(\frac{4\pi t}{T}\right)$, and therefore we obtain $W_3[U^\text{F}_{k_xk_yt}]=1$, for any value of $BT$ as long as the adiabatic limit holds.
\subsection{Doubling of the unit cell}\label{app:2UC}
\begin{figure}[b]
	\centering
	\includegraphics[width=\columnwidth]{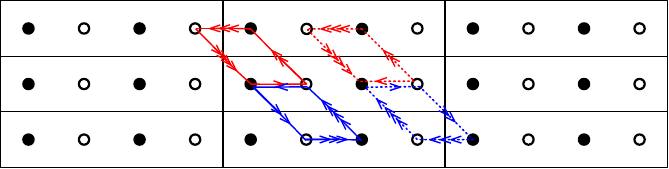}
	\caption{Redefinition of the unit cell. We consider a unit cell consisting of 4 orbitals. }
	\label{fig:breaktrans}
\end{figure}

We now consider a redefinition of the unit cell for the adiabatic $N=2$ Hopf pump of Sec.\ref{sec:N2Hopfp}. We double the unit cell in the $x$-direction, hence, there are 4 orbitals per unit cell, as shown on Fig.~\ref{fig:breaktrans}. The model has 2 bands, both doubly degenerate.  Thus the classification discussed in this article cannot apply in this example, as all the bands are not separated by a gap. 

We first consider the gauge $\ket{u_{k_xk_ytn}}$, $n=1,\dots,4$, that is obtained by Fourier transform of the following four WFs~(\ref{eq:WF1})-(\ref{eq:WF2}): $\ket{w_{R_xR_yt1}}$, $\ket{w_{R_xR_yt2}}$, $\ket{w_{(R_x+1)R_yt1}}$, and $\ket{w_{(R_x+1)R_yt2}}$, see Fig.~\ref{fig:breaktrans}. The direct calculation of the third winding number defined by such gauge choice gives $W_3[U_{k_xk_yt}]=1$.

Next, we introduce a change of basis between the two lower degenerate bands:
\begin{equation}
\begin{cases}
|\tilde{u}_{k_xk_yt1}\rangle=\alpha_{k_xk_yt}|u_{k_xk_yt1}\rangle +\beta_{k_xk_yt}|u_{k_xk_yt3}\rangle   \\
|\tilde{u}_{k_xk_yt3}\rangle=\gamma_{k_xk_yt}|u_{k_xk_yt1}\rangle +\delta_{k_xk_yt}|u_{k_xk_yt3}\rangle,
\end{cases}
\end{equation}\\
which defines a unitary $2\times 2$ matrix 
\begin{equation}
V_{k_xk_yt}=
\begin{pmatrix}
\alpha_{k_xk_yt} &\beta_{k_xk_yt}\\
\gamma_{k_xk_yt} &\delta_{k_xk_yt}
\end{pmatrix}.
\end{equation}
The above gauge defines a new unitary matrix $\widetilde{U}_{k_xk_yt}$. We make a choice for $V_{k_xk_yt}$ such that $W_3[V_{k_xk_yt}]=1$, this can be obtained by taking $V_{k_xk_yt}$ to be the unitary of the 2 sites unit cell example studied previously. Using this definition for the gauge transformation, we obtain $\widetilde{U}_{k_xk_yt}$
\begin{widetext}
\begin{equation}
\widetilde{U}_{k_xk_yt}=\begin{pmatrix}
\cos(t_0)^2 & \sin(t_0) &-\sin(t_0)\cos(t_0) &0 \\
-\sin(t_0)\cos(t_0) & \cos(t_0) &\sin(t_0)^2 & 0\\
 \cos(t_0)\sin(t_0) &0 &\cos(t_0)^2 & \sin(t_0)\\
  -\sin(t_0)^2  & 0&-\sin(t_0)\cos(t_0) &\cos(t_0) \\
\end{pmatrix},
\end{equation}
\begin{equation}
\widetilde{U}_{k_xk_yt}=\begin{pmatrix}
\sin(t_\frac{1}{4})^2e^{2ik_y} & \cos(t_\frac{1}{4}) &\sin(t_\frac{1}{4})\cos(t_{\frac{1}{4}})e^{ik_y} &0 \\
\cos(t_\frac{1}{4})\sin(t_\frac{1}{4})e^{ik_y}  & -\sin(t_\frac{1}{4})e^{-ik_y} &\cos(t_\frac{1}{4})^2 & 0\\
 -\sin(t_\frac{1}{4})\cos(t_\frac{1}{4})e^{ik_y}  &0 &\sin(t_\frac{1}{4})^2 &\cos(t_\frac{1}{4})\\
  -\cos(t_\frac{1}{4})^2  & 0&\cos(t_\frac{1}{4})\sin(t_{\frac{1}{4}})e^{-ik_y} &-\sin(t_\frac{1}{4})e^{-ik_y} \\
\end{pmatrix},
\end{equation}
\begin{equation}
\widetilde{U}_{k_xk_yt}=\begin{pmatrix}
\cos(t_\frac{1}{2})^2e^{2ik_y} & 0 &-\cos(t_\frac{1}{2})\sin(t_{\frac{1}{2}})e^{i(2k_y-k_x)}  &-\sin(t_\frac{1}{2})e^{-i(k_y-k_x)} \\
-\sin(t_\frac{1}{2})^2e^{ik_x}  & -\cos(t_\frac{1}{2})e^{-ik_y} &-\sin(t_\frac{1}{2})\cos(t_{\frac{1}{2}}) & 0\\
 \cos(t_\frac{1}{2})\sin(t_{\frac{1}{2}})e^{ik_x}   &-\sin(t_\frac{1}{2})e^{-ik_y} &\cos(t_\frac{1}{2})^2 &0\\
  -\sin(t_\frac{1}{2})\cos(t_{\frac{1}{2}})e^{i(2k_y-k_x)}  & 0&\sin(t_\frac{1}{2})^2e^{2i(k_y-k_x)} &-\cos(t_\frac{1}{2})e^{-ik_y} \\
\end{pmatrix},
\end{equation}
\begin{equation}
\widetilde{U}_{k_xk_yt}=\begin{pmatrix}
\sin(t_\frac{3}{4})^2 & 0 &\sin(t_\frac{3}{4})\cos(t_\frac{3}{4})e^{i(k_y-k_x)} &-\cos(t_\frac{3}{4})e^{-i(k_y-k_x)} \\
-\cos(t_\frac{3}{4})^2e^{ik_x}  & \sin(t_\frac{3}{4}) &\cos(t_\frac{3}{4})\sin(t_{\frac{3}{4}})e^{ik_y} & 0\\
 -\sin(t_\frac{3}{4})\cos(t_{\frac{3}{4}})e^{i(k_x-k_y)} &-\cos(t_\frac{3}{4})e^{-ik_y} &\sin(t_\frac{3}{4})^2 &0\\
  \cos(t_\frac{3}{4})\sin(t_\frac{3}{4})e^{i(k_y-k_x)}  & 0&\cos(t_\frac{3}{4})^2e^{2i(k_y-k_x)} &\sin(t_\frac{3}{4}) \\
\end{pmatrix},
\end{equation}
\end{widetext}
for the 4 different segments of the drive, where we introduced
	the notation $t_{n}=\frac{2\pi}{T}(t-nT)$.
We obtain that only the third and the fourth segments of the drive contribute to the third winding number, both giving an opposite $\frac{1}{2}$ contribution. Therefore $W_3[\widetilde{U}_{k_xk_yt}]=0$, demonstrating that the third winding number is not gauge independent in presence of the band degeneracies.

Furthermore, imposing open boundary conditions in the $y$-direction, the edge Chern number can be computed using the Wannier cut procedure described in the main text. We consider for concreteness 8 layers in the $y$-direction, which defines the ribbon supercell of 32 orbitals. We compute the bulk hybrid WFs $\ket{\tilde{w}_{k_xR_ytn}}$ from the Bloch eigenvectors $\ket{\tilde{u}_{k_xk_ytn}}$. Using these bulk hybrid WFs, we obtain the upper edge projector $\mathcal{P}_{k_xt}^{\text{edge}}$ by removing the 16 WFs from the bulk. Explicit computation leads to $\text{Ch}^\text{edge}=0$, since the contribution of the third and fourth time-segments cancel each other.

Lastly, we can define $(N=4)$-band Hopf pump using the Bloch eigenstates $\ket{\tilde{u}_{k_xk_ytn}}$ 
\begin{equation}
h_{k_xk_yt}=\sum_{n=1}^4 n \ket{\tilde{u}_{k_xk_ytn}}\bra{\tilde{u}_{k_xk_ytn}},
\label{eq:N4Hopfp}
\end{equation}
such that the 4 bands are now non-denegerate. The abelian part of the third Chern-Simons form can be computed explicitly, 
\begin{equation}
\sum_{n=1}^{4}P_3^n=-\frac{1}{6}.
\label{3rdcsnonzero}
\end{equation}
This Hamiltonian~(\ref{eq:N4Hopfp}) has $N_\text{Hopf}=0$, since the third winding number of $\widetilde{U}_{k_xk_yt}$ vanishes. For this example $N_\text{Hopf}\neq\sum_{n=1}^4 P_3^n$ holds, thus, the non-topological orbital magnetization $\sum_{n=1}^4 m_n^\text{nontop}=1/6$ does not vanish.

\subsection{The $N=3$ Hopf pump}\label{sec:3banddetails}
For $t\in \left[0,\frac{T}{6}\right]$, the states evolve as
\begin{equation}
\begin{cases}
	|u_{k_xk_yt1}\rangle=\cos\left(t_0\delta\right)|1\rangle-\sin\left(t_0\delta\right)e^{ik_x}|3\rangle,\\
|u_{k_xk_yt2}\rangle=|2\rangle,\\
|u_{k_xk_yt3}\rangle=\cos\left(t_0\delta\right)|3\rangle+\sin\left(t_0\delta\right)e^{-ik_x}|1\rangle,
\end{cases}
\end{equation}
where we introduced the notation $t_{n}=\frac{3\pi}{T}(t-nT)$ and the parameter $\delta\in]0,1[$ which creates the asymmetry
	between the trajectories of $|0\rangle$ and $|1\rangle$. For $t\in
	\left(\frac{T}{6},\frac{T}{3}\right]$, the states evolve as
\begin{equation}
\begin{cases}
	|u_{k_xk_yt1}\rangle=\cos(t_{\frac{1}{6}})\ket{u_{k_xk_y\frac{T}{6}1}}-\sin(t_{\frac{1}{6}})|2\rangle,\\
|u_{k_xk_yt2}\rangle=\sin(t_{\frac{1}{6}})\ket{u_{k_xk_y\frac{T}{6}1}}+\cos(t_{\frac{1}{6}})|2\rangle,\\
|u_{k_xk_yt3}\rangle=\ket{u_{k_xk_y\frac{T}{6}3}}.
\end{cases}
\end{equation}

For $t\in \left(\frac{T}{3},\frac{T}{2}\right]$, the states evolve as
\begin{equation}
\begin{cases}
	\ket{u_{k_xk_yt1}}=-\cos(t_{\frac{1}{3}})|2\rangle-\sin(t_{\frac{1}{3}})e^{-iK}\ket{u_{k_xk_y\frac{T}{6}1}},\\
\ket{u_{k_xk_yt2}}=\cos(t_{\frac{1}{3}})\ket{u_{k_xk_y\frac{T}{6}1}}-\sin(t_{\frac{1}{3}})e^{iK}|2\rangle,\\
\ket{u_{k_xk_yt3}}=\ket{u_{k_xk_y\frac{T}{6}1}},
\end{cases}
\end{equation}
where the notation $K=k_x+k_y$ has been introduced.
For $t\in \left(\frac{T}{2},\frac{2T}{3}\right]$, the states evolve as
\begin{equation}
\begin{cases}
\ket{u_{k_xk_yt1}}=-\cos\left(\delta(\frac{\pi}{2}-t_{\frac{1}{2}})\right)e^{-iK}|1\rangle\\
\quad+\sin\left(\delta(\frac{\pi}{2}-t_{\frac{1}{2}})\right)e^{-ik_y}|3\rangle,
\\
\ket{u_{k_xk_yt2}}=|2\left(\frac{T}{2}\right)\rangle,\\
\ket{u_{k_xk_yt3}}=\sin\left(\delta(\frac{\pi}{2}-t_{\frac{1}{2}})\right)e^{-ik_x}|1\rangle\\\quad +\cos\left(\delta(\frac{\pi}{2}-t_{\frac{1}{2}})\right)|3\rangle.
\end{cases}
\end{equation}
For $t\in \left(\frac{2T}{3},\frac{5T}{6}\right]$, the states evolve as
\begin{equation}
\begin{cases}
\ket{u_{k_xk_yt1}}=-\cos(t_{\frac{4}{6}})e^{-iK}|1\rangle+\sin(t_{\frac{4}{6}})e^{-ik_y}|2\rangle,\\
\ket{u_{k_xk_yt2}}=-\cos(t_{\frac{4}{6}})e^{iK}|2\rangle-\sin(t_{\frac{4}{6}})e^{ik_y}|1\rangle,\\
\ket{u_{k_xk_yt3}}=|3\rangle.
\end{cases}
\end{equation}
For $t\in \left(\frac{5T}{6},T\right)$, the states evolve as
\begin{equation}
\begin{cases}
\ket{u_{k_xk_yt1}}=\cos(t_{\frac{5}{6}})e^{-ik_y}|2\rangle+\sin(t_{\frac{5}{6}})|1\rangle,\\
\ket{u_{k_xk_yt2}}=-\cos(t_{\frac{5}{6}})e^{ik_y}|1\rangle+\sin(t_{\frac{5}{6}})|2\rangle,\\
\ket{u_{k_xk_yt3}}=|3\rangle.
\end{cases}
\end{equation}
\bigskip

\bibliography{refs}

\begin{thebibliography}{64}%
\makeatletter
\providecommand \@ifxundefined [1]{%
 \@ifx{#1\undefined}
}%
\providecommand \@ifnum [1]{%
 \ifnum #1\expandafter \@firstoftwo
 \else \expandafter \@secondoftwo
 \fi
}%
\providecommand \@ifx [1]{%
 \ifx #1\expandafter \@firstoftwo
 \else \expandafter \@secondoftwo
 \fi
}%
\providecommand \natexlab [1]{#1}%
\providecommand \enquote  [1]{``#1''}%
\providecommand \bibnamefont  [1]{#1}%
\providecommand \bibfnamefont [1]{#1}%
\providecommand \citenamefont [1]{#1}%
\providecommand \href@noop [0]{\@secondoftwo}%
\providecommand \href [0]{\begingroup \@sanitize@url \@href}%
\providecommand \@href[1]{\@@startlink{#1}\@@href}%
\providecommand \@@href[1]{\endgroup#1\@@endlink}%
\providecommand \@sanitize@url [0]{\catcode `\\12\catcode `\$12\catcode
  `\&12\catcode `\#12\catcode `\^12\catcode `\_12\catcode `\%12\relax}%
\providecommand \@@startlink[1]{}%
\providecommand \@@endlink[0]{}%
\providecommand \url  [0]{\begingroup\@sanitize@url \@url }%
\providecommand \@url [1]{\endgroup\@href {#1}{\urlprefix }}%
\providecommand \urlprefix  [0]{URL }%
\providecommand \Eprint [0]{\href }%
\providecommand \doibase [0]{http://dx.doi.org/}%
\providecommand \selectlanguage [0]{\@gobble}%
\providecommand \bibinfo  [0]{\@secondoftwo}%
\providecommand \bibfield  [0]{\@secondoftwo}%
\providecommand \translation [1]{[#1]}%
\providecommand \BibitemOpen [0]{}%
\providecommand \bibitemStop [0]{}%
\providecommand \bibitemNoStop [0]{.\EOS\space}%
\providecommand \EOS [0]{\spacefactor3000\relax}%
\providecommand \BibitemShut  [1]{\csname bibitem#1\endcsname}%
\let\auto@bib@innerbib\@empty
\bibitem [{\citenamefont {Chen}\ \emph {et~al.}(2011)\citenamefont {Chen},
  \citenamefont {Zhu}, \citenamefont {Xiao},\ and\ \citenamefont
  {Zhang}}]{chen2011}%
  \BibitemOpen
  \bibfield  {author} {\bibinfo {author} {\bibfnamefont {Hua}\ \bibnamefont
  {Chen}}, \bibinfo {author} {\bibfnamefont {Wenguang}\ \bibnamefont {Zhu}},
  \bibinfo {author} {\bibfnamefont {Di}~\bibnamefont {Xiao}}, \ and\ \bibinfo
  {author} {\bibfnamefont {Zhenyu}\ \bibnamefont {Zhang}},\ }\bibfield  {title}
  {\enquote {\bibinfo {title} {Co oxidation facilitated by robust surface
  states on au-covered topological insulators},}\ }\href {\doibase
  10.1103/PhysRevLett.107.056804} {\bibfield  {journal} {\bibinfo  {journal}
  {Phys. Rev. Lett.}\ }\textbf {\bibinfo {volume} {107}},\ \bibinfo {pages}
  {056804} (\bibinfo {year} {2011})}\BibitemShut {NoStop}%
\bibitem [{\citenamefont {Kitaev}(2001)}]{kitaev2001}%
  \BibitemOpen
  \bibfield  {author} {\bibinfo {author} {\bibfnamefont {A.~Yu.}\ \bibnamefont
  {Kitaev}},\ }\bibfield  {title} {\enquote {\bibinfo {title} {Unpaired
  majorana fermions in quantum wires},}\ }\href@noop {} {\bibfield  {journal}
  {\bibinfo  {journal} {Phys. Usp.}\ }\textbf {\bibinfo {volume} {44}},\
  \bibinfo {pages} {131} (\bibinfo {year} {2001})}\BibitemShut {NoStop}%
\bibitem [{\citenamefont {Kitaev}(2009)}]{kitaev2009}%
  \BibitemOpen
  \bibfield  {author} {\bibinfo {author} {\bibfnamefont {Alexei}\ \bibnamefont
  {Kitaev}},\ }\bibfield  {title} {\enquote {\bibinfo {title} {Periodic table
  for topological insulators and superconductors},}\ }\href {\doibase
  10.1063/1.3149495} {\bibfield  {journal} {\bibinfo  {journal} {AIP Conference
  Proceedings}\ }\textbf {\bibinfo {volume} {1134}},\ \bibinfo {pages} {22--30}
  (\bibinfo {year} {2009})}\BibitemShut {NoStop}%
\bibitem [{\citenamefont {Schnyder}\ \emph {et~al.}(2009)\citenamefont
  {Schnyder}, \citenamefont {Ryu}, \citenamefont {Furusaki},\ and\
  \citenamefont {Ludwig}}]{schnyder2009}%
  \BibitemOpen
  \bibfield  {author} {\bibinfo {author} {\bibfnamefont {Andreas~P.}\
  \bibnamefont {Schnyder}}, \bibinfo {author} {\bibfnamefont {Shinsei}\
  \bibnamefont {Ryu}}, \bibinfo {author} {\bibfnamefont {Akira}\ \bibnamefont
  {Furusaki}}, \ and\ \bibinfo {author} {\bibfnamefont {Andreas W.~W.}\
  \bibnamefont {Ludwig}},\ }\bibfield  {title} {\enquote {\bibinfo {title}
  {Classification of topological insulators and superconductors},}\ }\href
  {\doibase 10.1063/1.3149481} {\bibfield  {journal} {\bibinfo  {journal} {AIP
  Conference Proceedings}\ }\textbf {\bibinfo {volume} {1134}},\ \bibinfo
  {pages} {10--21} (\bibinfo {year} {2009})}\BibitemShut {NoStop}%
\bibitem [{\citenamefont {Turner}\ \emph {et~al.}(2012)\citenamefont {Turner},
  \citenamefont {Zhang}, \citenamefont {Mong},\ and\ \citenamefont
  {Vishwanath}}]{turner2012}%
  \BibitemOpen
  \bibfield  {author} {\bibinfo {author} {\bibfnamefont {Ari~M.}\ \bibnamefont
  {Turner}}, \bibinfo {author} {\bibfnamefont {Yi}~\bibnamefont {Zhang}},
  \bibinfo {author} {\bibfnamefont {Roger S.~K.}\ \bibnamefont {Mong}}, \ and\
  \bibinfo {author} {\bibfnamefont {Ashvin}\ \bibnamefont {Vishwanath}},\
  }\bibfield  {title} {\enquote {\bibinfo {title} {Quantized response and
  topology of magnetic insulators with inversion symmetry},}\ }\href {\doibase
  10.1103/PhysRevB.85.165120} {\bibfield  {journal} {\bibinfo  {journal} {Phys.
  Rev. B}\ }\textbf {\bibinfo {volume} {85}},\ \bibinfo {pages} {165120}
  (\bibinfo {year} {2012})}\BibitemShut {NoStop}%
\bibitem [{\citenamefont {Fu}(2011)}]{fu2011}%
  \BibitemOpen
  \bibfield  {author} {\bibinfo {author} {\bibfnamefont {Liang}\ \bibnamefont
  {Fu}},\ }\bibfield  {title} {\enquote {\bibinfo {title} {Topological
  crystalline insulators},}\ }\href {\doibase 10.1103/PhysRevLett.106.106802}
  {\bibfield  {journal} {\bibinfo  {journal} {Phys. Rev. Lett.}\ }\textbf
  {\bibinfo {volume} {106}},\ \bibinfo {pages} {106802} (\bibinfo {year}
  {2011})}\BibitemShut {NoStop}%
\bibitem [{\citenamefont {Trifunovic}\ and\ \citenamefont
  {Brouwer}(2017)}]{trifunovic2017}%
  \BibitemOpen
  \bibfield  {author} {\bibinfo {author} {\bibfnamefont {Luka}\ \bibnamefont
  {Trifunovic}}\ and\ \bibinfo {author} {\bibfnamefont {Piet~W.}\ \bibnamefont
  {Brouwer}},\ }\bibfield  {title} {\enquote {\bibinfo {title} {Bott
  periodicity for the topological classification of gapped states of matter
  with reflection symmetry},}\ }\href {\doibase 10.1103/PhysRevB.96.195109}
  {\bibfield  {journal} {\bibinfo  {journal} {Phys. Rev. B}\ }\textbf {\bibinfo
  {volume} {96}},\ \bibinfo {pages} {195109} (\bibinfo {year}
  {2017})}\BibitemShut {NoStop}%
\bibitem [{\citenamefont {Trifunovic}\ and\ \citenamefont
  {Brouwer}(2019)}]{trifunovic2019}%
  \BibitemOpen
  \bibfield  {author} {\bibinfo {author} {\bibfnamefont {Luka}\ \bibnamefont
  {Trifunovic}}\ and\ \bibinfo {author} {\bibfnamefont {Piet~W.}\ \bibnamefont
  {Brouwer}},\ }\bibfield  {title} {\enquote {\bibinfo {title} {Higher-order
  bulk-boundary correspondence for topological crystalline phases},}\ }\href
  {\doibase 10.1103/PhysRevX.9.011012} {\bibfield  {journal} {\bibinfo
  {journal} {Phys. Rev. X}\ }\textbf {\bibinfo {volume} {9}},\ \bibinfo {pages}
  {011012} (\bibinfo {year} {2019})}\BibitemShut {NoStop}%
\bibitem [{\citenamefont {Khalaf}\ \emph {et~al.}(2018)\citenamefont {Khalaf},
  \citenamefont {Po}, \citenamefont {Vishwanath},\ and\ \citenamefont
  {Watanabe}}]{khalaf2018}%
  \BibitemOpen
  \bibfield  {author} {\bibinfo {author} {\bibfnamefont {Eslam}\ \bibnamefont
  {Khalaf}}, \bibinfo {author} {\bibfnamefont {Hoi~Chun}\ \bibnamefont {Po}},
  \bibinfo {author} {\bibfnamefont {Ashvin}\ \bibnamefont {Vishwanath}}, \ and\
  \bibinfo {author} {\bibfnamefont {Haruki}\ \bibnamefont {Watanabe}},\
  }\bibfield  {title} {\enquote {\bibinfo {title} {Symmetry indicators and
  anomalous surface states of topological crystalline insulators},}\ }\href
  {\doibase 10.1103/PhysRevX.8.031070} {\bibfield  {journal} {\bibinfo
  {journal} {Phys. Rev. X}\ }\textbf {\bibinfo {volume} {8}},\ \bibinfo {pages}
  {031070} (\bibinfo {year} {2018})}\BibitemShut {NoStop}%
\bibitem [{\citenamefont {Bradlyn}\ \emph {et~al.}(2017)\citenamefont
  {Bradlyn}, \citenamefont {Elcoro}, \citenamefont {Cano}, \citenamefont
  {Vergniory}, \citenamefont {Wang}, \citenamefont {Felser}, \citenamefont
  {Aroyo},\ and\ \citenamefont {Bernevig}}]{bradlyn2017}%
  \BibitemOpen
  \bibfield  {author} {\bibinfo {author} {\bibfnamefont {Barry}\ \bibnamefont
  {Bradlyn}}, \bibinfo {author} {\bibfnamefont {L.}~\bibnamefont {Elcoro}},
  \bibinfo {author} {\bibfnamefont {Jennifer}\ \bibnamefont {Cano}}, \bibinfo
  {author} {\bibfnamefont {M.~G.}\ \bibnamefont {Vergniory}}, \bibinfo {author}
  {\bibfnamefont {Zhijun}\ \bibnamefont {Wang}}, \bibinfo {author}
  {\bibfnamefont {C.}~\bibnamefont {Felser}}, \bibinfo {author} {\bibfnamefont
  {M.~I.}\ \bibnamefont {Aroyo}}, \ and\ \bibinfo {author} {\bibfnamefont
  {B.~Andrei}\ \bibnamefont {Bernevig}},\ }\bibfield  {title} {\enquote
  {\bibinfo {title} {Topological quantum chemistry},}\ }\href@noop {}
  {\bibfield  {journal} {\bibinfo  {journal} {Nature}\ }\textbf {\bibinfo
  {volume} {547}},\ \bibinfo {pages} {298} (\bibinfo {year}
  {2017})}\BibitemShut {NoStop}%
\bibitem [{\citenamefont {Huang}\ \emph {et~al.}(2017)\citenamefont {Huang},
  \citenamefont {Song}, \citenamefont {Huang},\ and\ \citenamefont
  {Hermele}}]{huang2017}%
  \BibitemOpen
  \bibfield  {author} {\bibinfo {author} {\bibfnamefont {Sheng-Jie}\
  \bibnamefont {Huang}}, \bibinfo {author} {\bibfnamefont {Hao}\ \bibnamefont
  {Song}}, \bibinfo {author} {\bibfnamefont {Yi-Ping}\ \bibnamefont {Huang}}, \
  and\ \bibinfo {author} {\bibfnamefont {Michael}\ \bibnamefont {Hermele}},\
  }\bibfield  {title} {\enquote {\bibinfo {title} {Building crystalline
  topological phases from lower-dimensional states},}\ }\href {\doibase
  10.1103/PhysRevB.96.205106} {\bibfield  {journal} {\bibinfo  {journal} {Phys.
  Rev. B}\ }\textbf {\bibinfo {volume} {96}},\ \bibinfo {pages} {205106}
  (\bibinfo {year} {2017})}\BibitemShut {NoStop}%
\bibitem [{\citenamefont {Shiozaki}\ and\ \citenamefont
  {Sato}(2014)}]{shiozaki2014}%
  \BibitemOpen
  \bibfield  {author} {\bibinfo {author} {\bibfnamefont {Ken}\ \bibnamefont
  {Shiozaki}}\ and\ \bibinfo {author} {\bibfnamefont {Masatoshi}\ \bibnamefont
  {Sato}},\ }\bibfield  {title} {\enquote {\bibinfo {title} {Topology of
  crystalline insulators and superconductors},}\ }\href@noop {} {\bibfield
  {journal} {\bibinfo  {journal} {Phys. Rev. B}\ }\textbf {\bibinfo {volume}
  {90}},\ \bibinfo {pages} {165114} (\bibinfo {year} {2014})}\BibitemShut
  {NoStop}%
\bibitem [{\citenamefont {{Geier}}\ \emph {et~al.}(2019)\citenamefont
  {{Geier}}, \citenamefont {{Brouwer}},\ and\ \citenamefont
  {{Trifunovic}}}]{geier2019}%
  \BibitemOpen
  \bibfield  {author} {\bibinfo {author} {\bibfnamefont {Max}\ \bibnamefont
  {{Geier}}}, \bibinfo {author} {\bibfnamefont {Piet~W.}\ \bibnamefont
  {{Brouwer}}}, \ and\ \bibinfo {author} {\bibfnamefont {Luka}\ \bibnamefont
  {{Trifunovic}}},\ }\bibfield  {title} {\enquote {\bibinfo {title}
  {{Symmetry-based indicators for topological Bogoliubov-de Gennes
  Hamiltonians}},}\ }\href@noop {} {\bibfield  {journal} {\bibinfo  {journal}
  {arXiv:1910.11271}\ } (\bibinfo {year} {2019})}\BibitemShut {NoStop}%
\bibitem [{\citenamefont {{Ono}}\ \emph {et~al.}(2020)\citenamefont {{Ono}},
  \citenamefont {{Po}},\ and\ \citenamefont {{Shiozaki}}}]{ono2020}%
  \BibitemOpen
  \bibfield  {author} {\bibinfo {author} {\bibfnamefont {Seishiro}\
  \bibnamefont {{Ono}}}, \bibinfo {author} {\bibfnamefont {Hoi~Chun}\
  \bibnamefont {{Po}}}, \ and\ \bibinfo {author} {\bibfnamefont {Ken}\
  \bibnamefont {{Shiozaki}}},\ }\bibfield  {title} {\enquote {\bibinfo {title}
  {{$\mathbb{Z}_2$-enriched symmetry indicators for topological superconductors
  in the 1651 magnetic space groups}},}\ }\href@noop {} {\bibfield  {journal}
  {\bibinfo  {journal} {arXiv e-prints}\ ,\ \bibinfo {eid} {arXiv:2008.05499}}
  (\bibinfo {year} {2020})},\ \Eprint {http://arxiv.org/abs/2008.05499}
  {arXiv:2008.05499 [cond-mat.supr-con]} \BibitemShut {NoStop}%
\bibitem [{\citenamefont {Khalaf}(2018)}]{khalaf2018b}%
  \BibitemOpen
  \bibfield  {author} {\bibinfo {author} {\bibfnamefont {Eslam}\ \bibnamefont
  {Khalaf}},\ }\bibfield  {title} {\enquote {\bibinfo {title} {Higher-order
  topological insulators and superconductors protected by inversion
  symmetry},}\ }\href {\doibase 10.1103/PhysRevB.97.205136} {\bibfield
  {journal} {\bibinfo  {journal} {Phys. Rev. B}\ }\textbf {\bibinfo {volume}
  {97}},\ \bibinfo {pages} {205136} (\bibinfo {year} {2018})}\BibitemShut
  {NoStop}%
\bibitem [{\citenamefont {Schindler}\ \emph {et~al.}(2018)\citenamefont
  {Schindler}, \citenamefont {Cook}, \citenamefont {Vergniory}, \citenamefont
  {Wang}, \citenamefont {Parkin}, \citenamefont {Bernevig},\ and\ \citenamefont
  {Neupert}}]{schindler2018}%
  \BibitemOpen
  \bibfield  {author} {\bibinfo {author} {\bibfnamefont {Frank}\ \bibnamefont
  {Schindler}}, \bibinfo {author} {\bibfnamefont {Ashley~M.}\ \bibnamefont
  {Cook}}, \bibinfo {author} {\bibfnamefont {Maia~G.}\ \bibnamefont
  {Vergniory}}, \bibinfo {author} {\bibfnamefont {Zhijun}\ \bibnamefont
  {Wang}}, \bibinfo {author} {\bibfnamefont {Stuart S.~P.}\ \bibnamefont
  {Parkin}}, \bibinfo {author} {\bibfnamefont {B.~Andrei}\ \bibnamefont
  {Bernevig}}, \ and\ \bibinfo {author} {\bibfnamefont {Titus}\ \bibnamefont
  {Neupert}},\ }\bibfield  {title} {\enquote {\bibinfo {title} {Higher-order
  topological insulators},}\ }\href {\doibase 10.1126/sciadv.aat0346}
  {\bibfield  {journal} {\bibinfo  {journal} {Science Advances}\ }\textbf
  {\bibinfo {volume} {4}} (\bibinfo {year} {2018}),\
  10.1126/sciadv.aat0346}\BibitemShut {NoStop}%
\bibitem [{\citenamefont {Zhang}\ \emph {et~al.}(2019)\citenamefont {Zhang},
  \citenamefont {Jiang}, \citenamefont {Song}, \citenamefont {Huang},
  \citenamefont {He}, \citenamefont {Fang}, \citenamefont {Weng},\ and\
  \citenamefont {Fang}}]{zhang2019}%
  \BibitemOpen
  \bibfield  {author} {\bibinfo {author} {\bibfnamefont {Tiantian}\
  \bibnamefont {Zhang}}, \bibinfo {author} {\bibfnamefont {Yi}~\bibnamefont
  {Jiang}}, \bibinfo {author} {\bibfnamefont {Zhida}\ \bibnamefont {Song}},
  \bibinfo {author} {\bibfnamefont {He}~\bibnamefont {Huang}}, \bibinfo
  {author} {\bibfnamefont {Yuqing}\ \bibnamefont {He}}, \bibinfo {author}
  {\bibfnamefont {Zhong}\ \bibnamefont {Fang}}, \bibinfo {author}
  {\bibfnamefont {Hongming}\ \bibnamefont {Weng}}, \ and\ \bibinfo {author}
  {\bibfnamefont {Chen}\ \bibnamefont {Fang}},\ }\bibfield  {title} {\enquote
  {\bibinfo {title} {Catalogue of topological electronic materials},}\ }\href
  {\doibase 10.1038/s41586-019-0944-6} {\bibfield  {journal} {\bibinfo
  {journal} {Nature}\ }\textbf {\bibinfo {volume} {566}},\ \bibinfo {pages}
  {475} (\bibinfo {year} {2019})}\BibitemShut {NoStop}%
\bibitem [{\citenamefont {Trifunovic}\ and\ \citenamefont
  {Brouwer}(2021)}]{trifunovic2020a}%
  \BibitemOpen
  \bibfield  {author} {\bibinfo {author} {\bibfnamefont {Luka}\ \bibnamefont
  {Trifunovic}}\ and\ \bibinfo {author} {\bibfnamefont {Piet~W.}\ \bibnamefont
  {Brouwer}},\ }\bibfield  {title} {\enquote {\bibinfo {title} {Higher-order
  topological band structures},}\ }\href {\doibase
  https://doi.org/10.1002/pssb.202000090} {\bibfield  {journal} {\bibinfo
  {journal} {physica status solidi (b)}\ }\textbf {\bibinfo {volume} {258}},\
  \bibinfo {pages} {2000090} (\bibinfo {year} {2021})}\BibitemShut {NoStop}%
\bibitem [{Note1()}]{Note1}%
  \BibitemOpen
  \bibinfo {note} {In early studies~\cite {kennedy2014,kennedy2016} both
  delicate and fragile phases were called unstable topological phases, in order
  to distinguish them from the stable (tenfold-way) topological phases. In this
  work we borrow the terminology of Ref.~\protect \rev@citealpnum {nelson2020}
  and call phases {\protect \it delicate} if they are unstable but not
  fragile.}\BibitemShut {Stop}%
\bibitem [{\citenamefont {Moore}\ \emph {et~al.}(2008)\citenamefont {Moore},
  \citenamefont {Ran},\ and\ \citenamefont {Wen}}]{moore2008}%
  \BibitemOpen
  \bibfield  {author} {\bibinfo {author} {\bibfnamefont {Joel~E.}\ \bibnamefont
  {Moore}}, \bibinfo {author} {\bibfnamefont {Ying}\ \bibnamefont {Ran}}, \
  and\ \bibinfo {author} {\bibfnamefont {Xiao-Gang}\ \bibnamefont {Wen}},\
  }\bibfield  {title} {\enquote {\bibinfo {title} {Topological surface states
  in three-dimensional magnetic insulators},}\ }\href {\doibase
  10.1103/PhysRevLett.101.186805} {\bibfield  {journal} {\bibinfo  {journal}
  {Phys. Rev. Lett.}\ }\textbf {\bibinfo {volume} {101}},\ \bibinfo {pages}
  {186805} (\bibinfo {year} {2008})}\BibitemShut {NoStop}%
\bibitem [{\citenamefont {Kennedy}(2014)}]{kennedy2014}%
  \BibitemOpen
  \bibfield  {author} {\bibinfo {author} {\bibfnamefont {Ricardo}\ \bibnamefont
  {Kennedy}},\ }\emph {\bibinfo {title} {Homotopy Theory of Topological
  Insulators}},\ \href {https://kups.ub.uni-koeln.de/5873/} {Ph.D. thesis},\
  \bibinfo  {school} {Universit{\"a}t zu K{\"o}ln} (\bibinfo {year}
  {2014})\BibitemShut {NoStop}%
\bibitem [{\citenamefont {Po}\ \emph {et~al.}(2018)\citenamefont {Po},
  \citenamefont {Watanabe},\ and\ \citenamefont {Vishwanath}}]{po2018}%
  \BibitemOpen
  \bibfield  {author} {\bibinfo {author} {\bibfnamefont {Hoi~Chun}\
  \bibnamefont {Po}}, \bibinfo {author} {\bibfnamefont {Haruki}\ \bibnamefont
  {Watanabe}}, \ and\ \bibinfo {author} {\bibfnamefont {Ashvin}\ \bibnamefont
  {Vishwanath}},\ }\bibfield  {title} {\enquote {\bibinfo {title} {Fragile
  topology and wannier obstructions},}\ }\href {\doibase
  10.1103/PhysRevLett.121.126402} {\bibfield  {journal} {\bibinfo  {journal}
  {Phys. Rev. Lett.}\ }\textbf {\bibinfo {volume} {121}},\ \bibinfo {pages}
  {126402} (\bibinfo {year} {2018})}\BibitemShut {NoStop}%
\bibitem [{\citenamefont {Po}\ \emph {et~al.}(2017)\citenamefont {Po},
  \citenamefont {Vishwanath},\ and\ \citenamefont {Watanabe}}]{po2017}%
  \BibitemOpen
  \bibfield  {author} {\bibinfo {author} {\bibfnamefont {Hoi~Chun}\
  \bibnamefont {Po}}, \bibinfo {author} {\bibfnamefont {Ashvin}\ \bibnamefont
  {Vishwanath}}, \ and\ \bibinfo {author} {\bibfnamefont {Haruki}\ \bibnamefont
  {Watanabe}},\ }\bibfield  {title} {\enquote {\bibinfo {title} {Symmetry-based
  indicators of band topology in the 230 space groups},}\ }\href@noop {}
  {\bibfield  {journal} {\bibinfo  {journal} {Nature Comm.}\ }\textbf {\bibinfo
  {volume} {8}},\ \bibinfo {pages} {50} (\bibinfo {year} {2017})}\BibitemShut
  {NoStop}%
\bibitem [{\citenamefont {Benalcazar}\ \emph {et~al.}(2017)\citenamefont
  {Benalcazar}, \citenamefont {Bernevig},\ and\ \citenamefont
  {Hughes}}]{benalcazar2017}%
  \BibitemOpen
  \bibfield  {author} {\bibinfo {author} {\bibfnamefont {Wladimir~A.}\
  \bibnamefont {Benalcazar}}, \bibinfo {author} {\bibfnamefont {B.~Andrei}\
  \bibnamefont {Bernevig}}, \ and\ \bibinfo {author} {\bibfnamefont
  {Taylor~L.}\ \bibnamefont {Hughes}},\ }\bibfield  {title} {\enquote {\bibinfo
  {title} {Quantized electric multipole insulators},}\ }\href@noop {}
  {\bibfield  {journal} {\bibinfo  {journal} {Science}\ }\textbf {\bibinfo
  {volume} {357}},\ \bibinfo {pages} {61} (\bibinfo {year} {2017})}\BibitemShut
  {NoStop}%
\bibitem [{\citenamefont {Khalaf}\ \emph {et~al.}(2021)\citenamefont {Khalaf},
  \citenamefont {Benalcazar}, \citenamefont {Hughes},\ and\ \citenamefont
  {Queiroz}}]{khalaf2021}%
  \BibitemOpen
  \bibfield  {author} {\bibinfo {author} {\bibfnamefont {Eslam}\ \bibnamefont
  {Khalaf}}, \bibinfo {author} {\bibfnamefont {Wladimir~A.}\ \bibnamefont
  {Benalcazar}}, \bibinfo {author} {\bibfnamefont {Taylor~L.}\ \bibnamefont
  {Hughes}}, \ and\ \bibinfo {author} {\bibfnamefont {Raquel}\ \bibnamefont
  {Queiroz}},\ }\bibfield  {title} {\enquote {\bibinfo {title}
  {Boundary-obstructed topological phases},}\ }\href {\doibase
  10.1103/PhysRevResearch.3.013239} {\bibfield  {journal} {\bibinfo  {journal}
  {Phys. Rev. Research}\ }\textbf {\bibinfo {volume} {3}},\ \bibinfo {pages}
  {013239} (\bibinfo {year} {2021})}\BibitemShut {NoStop}%
\bibitem [{Note2()}]{Note2}%
  \BibitemOpen
  \bibinfo {note} {The Wannier gap, unlike the band gap, is not physical
  observable; It is unclear how to define it for interacting
  systems.}\BibitemShut {Stop}%
\bibitem [{\citenamefont {Song}\ \emph {et~al.}(2020)\citenamefont {Song},
  \citenamefont {Elcoro},\ and\ \citenamefont {Bernevig}}]{song2020}%
  \BibitemOpen
  \bibfield  {author} {\bibinfo {author} {\bibfnamefont {Zhi-Da}\ \bibnamefont
  {Song}}, \bibinfo {author} {\bibfnamefont {Luis}\ \bibnamefont {Elcoro}}, \
  and\ \bibinfo {author} {\bibfnamefont {B.~Andrei}\ \bibnamefont {Bernevig}},\
  }\bibfield  {title} {\enquote {\bibinfo {title} {Twisted bulk-boundary
  correspondence of fragile topology},}\ }\href {\doibase
  10.1126/science.aaz7650} {\bibfield  {journal} {\bibinfo  {journal}
  {Science}\ }\textbf {\bibinfo {volume} {367}},\ \bibinfo {pages} {794--797}
  (\bibinfo {year} {2020})}\BibitemShut {NoStop}%
\bibitem [{\citenamefont {Alexandradinata}\ \emph {et~al.}(2021)\citenamefont
  {Alexandradinata}, \citenamefont {Nelson},\ and\ \citenamefont
  {Soluyanov}}]{alexandradinata2020}%
  \BibitemOpen
  \bibfield  {author} {\bibinfo {author} {\bibfnamefont {A.}~\bibnamefont
  {Alexandradinata}}, \bibinfo {author} {\bibfnamefont {Aleksandra}\
  \bibnamefont {Nelson}}, \ and\ \bibinfo {author} {\bibfnamefont {Alexey~A.}\
  \bibnamefont {Soluyanov}},\ }\bibfield  {title} {\enquote {\bibinfo {title}
  {Teleportation of berry curvature on the surface of a hopf insulator},}\
  }\href {\doibase 10.1103/PhysRevB.103.045107} {\bibfield  {journal} {\bibinfo
   {journal} {Phys. Rev. B}\ }\textbf {\bibinfo {volume} {103}},\ \bibinfo
  {pages} {045107} (\bibinfo {year} {2021})}\BibitemShut {NoStop}%
\bibitem [{\citenamefont {Liu}\ \emph {et~al.}(2017)\citenamefont {Liu},
  \citenamefont {Vafa},\ and\ \citenamefont {Xu}}]{liu2017}%
  \BibitemOpen
  \bibfield  {author} {\bibinfo {author} {\bibfnamefont {Chunxiao}\
  \bibnamefont {Liu}}, \bibinfo {author} {\bibfnamefont {Farzan}\ \bibnamefont
  {Vafa}}, \ and\ \bibinfo {author} {\bibfnamefont {Cenke}\ \bibnamefont
  {Xu}},\ }\bibfield  {title} {\enquote {\bibinfo {title} {Symmetry-protected
  topological hopf insulator and its generalizations},}\ }\href {\doibase
  10.1103/PhysRevB.95.161116} {\bibfield  {journal} {\bibinfo  {journal} {Phys.
  Rev. B}\ }\textbf {\bibinfo {volume} {95}},\ \bibinfo {pages} {161116}
  (\bibinfo {year} {2017})}\BibitemShut {NoStop}%
\bibitem [{\citenamefont {Roy}\ and\ \citenamefont {Harper}(2017)}]{roy2017}%
  \BibitemOpen
  \bibfield  {author} {\bibinfo {author} {\bibfnamefont {Rahul}\ \bibnamefont
  {Roy}}\ and\ \bibinfo {author} {\bibfnamefont {Fenner}\ \bibnamefont
  {Harper}},\ }\bibfield  {title} {\enquote {\bibinfo {title} {Periodic table
  for floquet topological insulators},}\ }\href {\doibase
  10.1103/PhysRevB.96.155118} {\bibfield  {journal} {\bibinfo  {journal} {Phys.
  Rev. B}\ }\textbf {\bibinfo {volume} {96}},\ \bibinfo {pages} {155118}
  (\bibinfo {year} {2017})}\BibitemShut {NoStop}%
\bibitem [{\citenamefont {Ahn}\ \emph {et~al.}(2018)\citenamefont {Ahn},
  \citenamefont {Kim}, \citenamefont {Kim},\ and\ \citenamefont
  {Yang}}]{ahn2018}%
  \BibitemOpen
  \bibfield  {author} {\bibinfo {author} {\bibfnamefont {Junyeong}\
  \bibnamefont {Ahn}}, \bibinfo {author} {\bibfnamefont {Dongwook}\
  \bibnamefont {Kim}}, \bibinfo {author} {\bibfnamefont {Youngkuk}\
  \bibnamefont {Kim}}, \ and\ \bibinfo {author} {\bibfnamefont {Bohm-Jung}\
  \bibnamefont {Yang}},\ }\bibfield  {title} {\enquote {\bibinfo {title} {Band
  topology and linking structure of nodal line semimetals with ${Z}_{2}$
  monopole charges},}\ }\href {\doibase 10.1103/PhysRevLett.121.106403}
  {\bibfield  {journal} {\bibinfo  {journal} {Phys. Rev. Lett.}\ }\textbf
  {\bibinfo {volume} {121}},\ \bibinfo {pages} {106403} (\bibinfo {year}
  {2018})}\BibitemShut {NoStop}%
\bibitem [{\citenamefont {Wu}\ \emph {et~al.}(2019)\citenamefont {Wu},
  \citenamefont {Soluyanov},\ and\ \citenamefont {Bzdu{\v s}ek}}]{wu2019}%
  \BibitemOpen
  \bibfield  {author} {\bibinfo {author} {\bibfnamefont {QuanSheng}\
  \bibnamefont {Wu}}, \bibinfo {author} {\bibfnamefont {Alexey~A.}\
  \bibnamefont {Soluyanov}}, \ and\ \bibinfo {author} {\bibfnamefont
  {Tom{\'a}{\v s}}\ \bibnamefont {Bzdu{\v s}ek}},\ }\bibfield  {title}
  {\enquote {\bibinfo {title} {Non-abelian band topology in noninteracting
  metals},}\ }\href {\doibase 10.1126/science.aau8740} {\bibfield  {journal}
  {\bibinfo  {journal} {Science}\ }\textbf {\bibinfo {volume} {365}},\ \bibinfo
  {pages} {1273--1277} (\bibinfo {year} {2019})}\BibitemShut {NoStop}%
\bibitem [{\citenamefont {Tiwari}\ and\ \citenamefont
  {Bzdu\ifmmode~\check{s}\else \v{s}\fi{}ek}(2020)}]{apoorv2020}%
  \BibitemOpen
  \bibfield  {author} {\bibinfo {author} {\bibfnamefont {Apoorv}\ \bibnamefont
  {Tiwari}}\ and\ \bibinfo {author} {\bibfnamefont {Tom\'a\ifmmode
  \check{s}\else~\v{s}\fi{}}\ \bibnamefont {Bzdu\ifmmode~\check{s}\else
  \v{s}\fi{}ek}},\ }\bibfield  {title} {\enquote {\bibinfo {title} {Non-abelian
  topology of nodal-line rings in $\mathcal{PT}$-symmetric systems},}\ }\href
  {\doibase 10.1103/PhysRevB.101.195130} {\bibfield  {journal} {\bibinfo
  {journal} {Phys. Rev. B}\ }\textbf {\bibinfo {volume} {101}},\ \bibinfo
  {pages} {195130} (\bibinfo {year} {2020})}\BibitemShut {NoStop}%
\bibitem [{Note3()}]{Note3}%
  \BibitemOpen
  \bibinfo {note} {The magnetoelectric polarizability tensor is isotropic if
  the contributions from all the bands are considered, see Ref~\protect
  \rev@citealpnum {essin2010}.}\BibitemShut {Stop}%
\bibitem [{\citenamefont {Thouless}(1983)}]{thouless1983}%
  \BibitemOpen
  \bibfield  {author} {\bibinfo {author} {\bibfnamefont {D.~J.}\ \bibnamefont
  {Thouless}},\ }\bibfield  {title} {\enquote {\bibinfo {title} {Quantization
  of particle transport},}\ }\href {\doibase 10.1103/PhysRevB.27.6083}
  {\bibfield  {journal} {\bibinfo  {journal} {Phys. Rev. B}\ }\textbf {\bibinfo
  {volume} {27}},\ \bibinfo {pages} {6083--6087} (\bibinfo {year}
  {1983})}\BibitemShut {NoStop}%
\bibitem [{\citenamefont {Rudner}\ \emph {et~al.}(2013)\citenamefont {Rudner},
  \citenamefont {Lindner}, \citenamefont {Berg},\ and\ \citenamefont
  {Levin}}]{rudner2013}%
  \BibitemOpen
  \bibfield  {author} {\bibinfo {author} {\bibfnamefont {Mark~S.}\ \bibnamefont
  {Rudner}}, \bibinfo {author} {\bibfnamefont {Netanel~H.}\ \bibnamefont
  {Lindner}}, \bibinfo {author} {\bibfnamefont {Erez}\ \bibnamefont {Berg}}, \
  and\ \bibinfo {author} {\bibfnamefont {Michael}\ \bibnamefont {Levin}},\
  }\bibfield  {title} {\enquote {\bibinfo {title} {Anomalous edge states and
  the bulk-edge correspondence for periodically driven two-dimensional
  systems},}\ }\href {\doibase 10.1103/PhysRevX.3.031005} {\bibfield  {journal}
  {\bibinfo  {journal} {Phys. Rev. X}\ }\textbf {\bibinfo {volume} {3}},\
  \bibinfo {pages} {031005} (\bibinfo {year} {2013})}\BibitemShut {NoStop}%
\bibitem [{\citenamefont {Pontryagin}(1941)}]{pontryagin1941}%
  \BibitemOpen
  \bibfield  {author} {\bibinfo {author} {\bibfnamefont {L.}~\bibnamefont
  {Pontryagin}},\ }\bibfield  {title} {\enquote {\bibinfo {title} {A
  classification of mappings of the three-dimensional complex into the
  two-dimensional sphere},}\ }\href@noop {} {\bibfield  {journal} {\bibinfo
  {journal} {Rec. Math.}\ }\textbf {\bibinfo {volume} {9}},\ \bibinfo {pages}
  {331} (\bibinfo {year} {1941})}\BibitemShut {NoStop}%
\bibitem [{Note4()}]{Note4}%
  \BibitemOpen
  \bibinfo {note} {For the tenfold-way classification, the distinction between
  strong and weak topological invariants is related to the effects of disorder;
  The value of a strong topological invariant cannot change due to inclusion of
  translation-symmetry-breaking perturbations. On the other hand, delicate
  topological phases depend crucially on the presence of translational
  symmetry. Hence, in this work we use a more general definition, where the
  strong topological invariants are those invariants that can be defined on a
  $d$-dimensional sphere instead of BZ.}\BibitemShut {Stop}%
\bibitem [{\citenamefont {Hatcher}(2003)}]{hatcher}%
  \BibitemOpen
  \bibfield  {author} {\bibinfo {author} {\bibfnamefont {A.}~\bibnamefont
  {Hatcher}},\ }\href@noop {} {\emph {\bibinfo {title} {Vector Bundles and
  K-Theory}}}\ (\bibinfo {year} {2003})\BibitemShut {NoStop}%
\bibitem [{\citenamefont {\"Unal}\ \emph {et~al.}(2019)\citenamefont {\"Unal},
  \citenamefont {Eckardt},\ and\ \citenamefont {Slager}}]{unal2019}%
  \BibitemOpen
  \bibfield  {author} {\bibinfo {author} {\bibfnamefont {F.~Nur}\ \bibnamefont
  {\"Unal}}, \bibinfo {author} {\bibfnamefont {Andr\'e}\ \bibnamefont
  {Eckardt}}, \ and\ \bibinfo {author} {\bibfnamefont {Robert-Jan}\
  \bibnamefont {Slager}},\ }\bibfield  {title} {\enquote {\bibinfo {title}
  {Hopf characterization of two-dimensional floquet topological insulators},}\
  }\href {\doibase 10.1103/PhysRevResearch.1.022003} {\bibfield  {journal}
  {\bibinfo  {journal} {Phys. Rev. Research}\ }\textbf {\bibinfo {volume}
  {1}},\ \bibinfo {pages} {022003} (\bibinfo {year} {2019})}\BibitemShut
  {NoStop}%
\bibitem [{Note5()}]{Note5}%
  \BibitemOpen
  \bibinfo {note} {There are more possibilities herein,~\cite {bouhon2020b} one
  can define fragile classifications by allowing only certain ranks ${\protect
  \cal P}_{\protect \vec k}^n$ to be varied, although the physical relevance of
  such classification schemes is unclear.}\BibitemShut {Stop}%
\bibitem [{\citenamefont {Bouhon}\ \emph
  {et~al.}(2020{\natexlab{a}})\citenamefont {Bouhon}, \citenamefont {Wu},
  \citenamefont {Slager}, \citenamefont {Weng}, \citenamefont {Yazyev},\ and\
  \citenamefont {Bzdu{\v{s}}ek}}]{bouhon2020}%
  \BibitemOpen
  \bibfield  {author} {\bibinfo {author} {\bibfnamefont {Adrien}\ \bibnamefont
  {Bouhon}}, \bibinfo {author} {\bibfnamefont {QuanSheng}\ \bibnamefont {Wu}},
  \bibinfo {author} {\bibfnamefont {Robert-Jan}\ \bibnamefont {Slager}},
  \bibinfo {author} {\bibfnamefont {Hongming}\ \bibnamefont {Weng}}, \bibinfo
  {author} {\bibfnamefont {Oleg~V.}\ \bibnamefont {Yazyev}}, \ and\ \bibinfo
  {author} {\bibfnamefont {Tom{\'a}{\v{s}}}\ \bibnamefont {Bzdu{\v{s}}ek}},\
  }\bibfield  {title} {\enquote {\bibinfo {title} {Non-abelian reciprocal
  braiding of weyl points and its manifestation in zrte},}\ }\href {\doibase
  10.1038/s41567-020-0967-9} {\bibfield  {journal} {\bibinfo  {journal} {Nature
  Physics}\ }\textbf {\bibinfo {volume} {16}},\ \bibinfo {pages} {1137--1143}
  (\bibinfo {year} {2020}{\natexlab{a}})}\BibitemShut {NoStop}%
\bibitem [{Note6()}]{Note6}%
  \BibitemOpen
  \bibinfo {note} {The group $\pi _i(X,A)$ is not isomorphic to $\pi _i(X/A)$
  in general. For example, when $X=D^2$ and $A=S^1$, the group $\pi
  _i(D^2,S^2)$ is trivial for $i>2$ as seen by exact sequence similar to
  Eq.~(\ref {eq:3}), whereas the $\pi _i(D^2/S^1=S^2)$ is non-trivial for
  infinitely many values of $i$. The isomorphism~(\ref {eq:9}) follows directly
  from the long exact sequence for the fibration $U(1)^N\rightarrow
  U(N)\rightarrow U(N)/U(1)^N$.}\BibitemShut {Stop}%
\bibitem [{\citenamefont {Ryu}\ \emph {et~al.}(2010)\citenamefont {Ryu},
  \citenamefont {Schnyder}, \citenamefont {Furusaki},\ and\ \citenamefont
  {Ludwig}}]{ryu2010}%
  \BibitemOpen
  \bibfield  {author} {\bibinfo {author} {\bibfnamefont {Shinsei}\ \bibnamefont
  {Ryu}}, \bibinfo {author} {\bibfnamefont {Andreas~P}\ \bibnamefont
  {Schnyder}}, \bibinfo {author} {\bibfnamefont {Akira}\ \bibnamefont
  {Furusaki}}, \ and\ \bibinfo {author} {\bibfnamefont {Andreas W~W}\
  \bibnamefont {Ludwig}},\ }\bibfield  {title} {\enquote {\bibinfo {title}
  {Topological insulators and superconductors: tenfold way and dimensional
  hierarchy},}\ }\href@noop {} {\bibfield  {journal} {\bibinfo  {journal} {New
  Journal of Physics}\ }\textbf {\bibinfo {volume} {12}},\ \bibinfo {pages}
  {065010} (\bibinfo {year} {2010})}\BibitemShut {NoStop}%
\bibitem [{\citenamefont {Qi}\ and\ \citenamefont {Zhang}(2008)}]{qi2008}%
  \BibitemOpen
  \bibfield  {author} {\bibinfo {author} {\bibfnamefont {Xiao-Liang}\
  \bibnamefont {Qi}}\ and\ \bibinfo {author} {\bibfnamefont {Shou-Cheng}\
  \bibnamefont {Zhang}},\ }\bibfield  {title} {\enquote {\bibinfo {title}
  {Spin-charge separation in the quantum spin hall state},}\ }\href {\doibase
  10.1103/PhysRevLett.101.086802} {\bibfield  {journal} {\bibinfo  {journal}
  {Phys. Rev. Lett.}\ }\textbf {\bibinfo {volume} {101}},\ \bibinfo {pages}
  {086802} (\bibinfo {year} {2008})}\BibitemShut {NoStop}%
\bibitem [{\citenamefont {Essin}\ \emph {et~al.}(2010)\citenamefont {Essin},
  \citenamefont {Turner}, \citenamefont {Moore},\ and\ \citenamefont
  {Vanderbilt}}]{essin2010}%
  \BibitemOpen
  \bibfield  {author} {\bibinfo {author} {\bibfnamefont {Andrew~M.}\
  \bibnamefont {Essin}}, \bibinfo {author} {\bibfnamefont {Ari~M.}\
  \bibnamefont {Turner}}, \bibinfo {author} {\bibfnamefont {Joel~E.}\
  \bibnamefont {Moore}}, \ and\ \bibinfo {author} {\bibfnamefont {David}\
  \bibnamefont {Vanderbilt}},\ }\bibfield  {title} {\enquote {\bibinfo {title}
  {Orbital magnetoelectric coupling in band insulators},}\ }\href {\doibase
  10.1103/PhysRevB.81.205104} {\bibfield  {journal} {\bibinfo  {journal} {Phys.
  Rev. B}\ }\textbf {\bibinfo {volume} {81}},\ \bibinfo {pages} {205104}
  (\bibinfo {year} {2010})}\BibitemShut {NoStop}%
\bibitem [{\citenamefont {Trifunovic}(2020)}]{trifunovic2020}%
  \BibitemOpen
  \bibfield  {author} {\bibinfo {author} {\bibfnamefont {Luka}\ \bibnamefont
  {Trifunovic}},\ }\bibfield  {title} {\enquote {\bibinfo {title}
  {Bulk-and-edge to corner correspondence},}\ }\href {\doibase
  10.1103/PhysRevResearch.2.043012} {\bibfield  {journal} {\bibinfo  {journal}
  {Phys. Rev. Research}\ }\textbf {\bibinfo {volume} {2}},\ \bibinfo {pages}
  {043012} (\bibinfo {year} {2020})}\BibitemShut {NoStop}%
\bibitem [{Note7()}]{Note7}%
  \BibitemOpen
  \bibinfo {note} {Since the Wannier cut is performed on all the bands, unlike
  in Ref.~\protect \rev@citealpnum {trifunovic2020}, no condition on the
  crystal's termination needs to be imposed, i.e., a metallic termination is
  allowed.}\BibitemShut {Stop}%
\bibitem [{\citenamefont {Olsen}\ \emph {et~al.}(2017)\citenamefont {Olsen},
  \citenamefont {Taherinejad}, \citenamefont {Vanderbilt},\ and\ \citenamefont
  {Souza}}]{olsen2017}%
  \BibitemOpen
  \bibfield  {author} {\bibinfo {author} {\bibfnamefont {Thomas}\ \bibnamefont
  {Olsen}}, \bibinfo {author} {\bibfnamefont {Maryam}\ \bibnamefont
  {Taherinejad}}, \bibinfo {author} {\bibfnamefont {David}\ \bibnamefont
  {Vanderbilt}}, \ and\ \bibinfo {author} {\bibfnamefont {Ivo}\ \bibnamefont
  {Souza}},\ }\bibfield  {title} {\enquote {\bibinfo {title} {Surface theorem
  for the chern-simons axion coupling},}\ }\href {\doibase
  10.1103/PhysRevB.95.075137} {\bibfield  {journal} {\bibinfo  {journal} {Phys.
  Rev. B}\ }\textbf {\bibinfo {volume} {95}},\ \bibinfo {pages} {075137}
  (\bibinfo {year} {2017})}\BibitemShut {NoStop}%
\bibitem [{\citenamefont {Zhu}\ \emph {et~al.}(2021)\citenamefont {Zhu},
  \citenamefont {Hughes},\ and\ \citenamefont {Alexandradinata}}]{zhu2021}%
  \BibitemOpen
  \bibfield  {author} {\bibinfo {author} {\bibfnamefont {Penghao}\ \bibnamefont
  {Zhu}}, \bibinfo {author} {\bibfnamefont {Taylor~L.}\ \bibnamefont {Hughes}},
  \ and\ \bibinfo {author} {\bibfnamefont {A.}~\bibnamefont
  {Alexandradinata}},\ }\bibfield  {title} {\enquote {\bibinfo {title}
  {Quantized surface magnetism and higher-order topology: Application to the
  hopf insulator},}\ }\href {\doibase 10.1103/PhysRevB.103.014417} {\bibfield
  {journal} {\bibinfo  {journal} {Phys. Rev. B}\ }\textbf {\bibinfo {volume}
  {103}},\ \bibinfo {pages} {014417} (\bibinfo {year} {2021})}\BibitemShut
  {NoStop}%
\bibitem [{Note8()}]{Note8}%
  \BibitemOpen
  \bibinfo {note} {A direct consequence of this non-uniqueness is inability to
  uniquely define edge polarization and quadrupole moment of two-dimensional
  insulators~\cite {trifunovic2020,ren2021}}\BibitemShut {NoStop}%
\bibitem [{\citenamefont {{Nelson}}\ \emph {et~al.}(2020)\citenamefont
  {{Nelson}}, \citenamefont {{Neupert}}, \citenamefont {{Bzdu{\v{s}}ek}},\ and\
  \citenamefont {{Alexandradinata}}}]{nelson2020}%
  \BibitemOpen
  \bibfield  {author} {\bibinfo {author} {\bibfnamefont {Aleksandra}\
  \bibnamefont {{Nelson}}}, \bibinfo {author} {\bibfnamefont {Titus}\
  \bibnamefont {{Neupert}}}, \bibinfo {author} {\bibfnamefont
  {Tom{\'a}{\v{s}}}\ \bibnamefont {{Bzdu{\v{s}}ek}}}, \ and\ \bibinfo {author}
  {\bibfnamefont {A.}~\bibnamefont {{Alexandradinata}}},\ }\bibfield  {title}
  {\enquote {\bibinfo {title} {{Multicellularity of delicate topological
  insulators}},}\ }\href@noop {} {\bibfield  {journal} {\bibinfo  {journal}
  {arXiv e-prints}\ ,\ \bibinfo {eid} {arXiv:2009.01863}} (\bibinfo {year}
  {2020})},\ \Eprint {http://arxiv.org/abs/2009.01863} {arXiv:2009.01863
  [cond-mat.mes-hall]} \BibitemShut {NoStop}%
\bibitem [{\citenamefont {Streda}(1982)}]{streda1982}%
  \BibitemOpen
  \bibfield  {author} {\bibinfo {author} {\bibfnamefont {P}~\bibnamefont
  {Streda}},\ }\bibfield  {title} {\enquote {\bibinfo {title} {Theory of
  quantised hall conductivity in two dimensions},}\ }\href {\doibase
  10.1088/0022-3719/15/22/005} {\bibfield  {journal} {\bibinfo  {journal}
  {Journal of Physics C: Solid State Physics}\ }\textbf {\bibinfo {volume}
  {15}},\ \bibinfo {pages} {L717--L721} (\bibinfo {year} {1982})}\BibitemShut
  {NoStop}%
\bibitem [{Note9()}]{Note9}%
  \BibitemOpen
  \bibinfo {note} {The more precise statement is that the total shift from all
  the edge bands is $N_\protect \text {Hopf}$, i.e., the shift does not need to
  be carried by a single band.}\BibitemShut {Stop}%
\bibitem [{\citenamefont {Trifunovic}\ \emph {et~al.}(2019)\citenamefont
  {Trifunovic}, \citenamefont {Ono},\ and\ \citenamefont
  {Watanabe}}]{trifunovic2019a}%
  \BibitemOpen
  \bibfield  {author} {\bibinfo {author} {\bibfnamefont {Luka}\ \bibnamefont
  {Trifunovic}}, \bibinfo {author} {\bibfnamefont {Seishiro}\ \bibnamefont
  {Ono}}, \ and\ \bibinfo {author} {\bibfnamefont {Haruki}\ \bibnamefont
  {Watanabe}},\ }\bibfield  {title} {\enquote {\bibinfo {title} {Geometric
  orbital magnetization in adiabatic processes},}\ }\href {\doibase
  10.1103/PhysRevB.100.054408} {\bibfield  {journal} {\bibinfo  {journal}
  {Phys. Rev. B}\ }\textbf {\bibinfo {volume} {100}},\ \bibinfo {pages}
  {054408} (\bibinfo {year} {2019})}\BibitemShut {NoStop}%
\bibitem [{\citenamefont {Thonhauser}\ \emph {et~al.}(2005)\citenamefont
  {Thonhauser}, \citenamefont {Ceresoli}, \citenamefont {Vanderbilt},\ and\
  \citenamefont {Resta}}]{thonhauser2005}%
  \BibitemOpen
  \bibfield  {author} {\bibinfo {author} {\bibfnamefont {T.}~\bibnamefont
  {Thonhauser}}, \bibinfo {author} {\bibfnamefont {Davide}\ \bibnamefont
  {Ceresoli}}, \bibinfo {author} {\bibfnamefont {David}\ \bibnamefont
  {Vanderbilt}}, \ and\ \bibinfo {author} {\bibfnamefont {R.}~\bibnamefont
  {Resta}},\ }\bibfield  {title} {\enquote {\bibinfo {title} {Orbital
  magnetization in periodic insulators},}\ }\href {\doibase
  10.1103/PhysRevLett.95.137205} {\bibfield  {journal} {\bibinfo  {journal}
  {Phys. Rev. Lett.}\ }\textbf {\bibinfo {volume} {95}},\ \bibinfo {pages}
  {137205} (\bibinfo {year} {2005})}\BibitemShut {NoStop}%
\bibitem [{Note10()}]{Note10}%
  \BibitemOpen
  \bibinfo {note} {Such ``leakage'' occurs also for time-independent band
  insulators””~\cite {ren2021}}\BibitemShut {NoStop}%
\bibitem [{\citenamefont {Berry}(1984)}]{berry1984}%
  \BibitemOpen
  \bibfield  {author} {\bibinfo {author} {\bibfnamefont {Michael~Victor}\
  \bibnamefont {Berry}},\ }\bibfield  {title} {\enquote {\bibinfo {title}
  {Quantal phase factors accompanying adiabatic changes},}\ }\href {\doibase
  10.1098/rspa.1984.0023} {\bibfield  {journal} {\bibinfo  {journal}
  {Proceedings of the Royal Society of London. A. Mathematical and Physical
  Sciences}\ }\textbf {\bibinfo {volume} {392}},\ \bibinfo {pages} {45--57}
  (\bibinfo {year} {1984})}\BibitemShut {NoStop}%
\bibitem [{\citenamefont {Ren}\ \emph {et~al.}(2021)\citenamefont {Ren},
  \citenamefont {Souza},\ and\ \citenamefont {Vanderbilt}}]{ren2021}%
  \BibitemOpen
  \bibfield  {author} {\bibinfo {author} {\bibfnamefont {Shang}\ \bibnamefont
  {Ren}}, \bibinfo {author} {\bibfnamefont {Ivo}\ \bibnamefont {Souza}}, \ and\
  \bibinfo {author} {\bibfnamefont {David}\ \bibnamefont {Vanderbilt}},\
  }\bibfield  {title} {\enquote {\bibinfo {title} {Quadrupole moments, edge
  polarizations, and corner charges in the wannier representation},}\ }\href
  {\doibase 10.1103/PhysRevB.103.035147} {\bibfield  {journal} {\bibinfo
  {journal} {Phys. Rev. B}\ }\textbf {\bibinfo {volume} {103}},\ \bibinfo
  {pages} {035147} (\bibinfo {year} {2021})}\BibitemShut {NoStop}%
\bibitem [{\citenamefont {Titum}\ \emph {et~al.}(2016)\citenamefont {Titum},
  \citenamefont {Berg}, \citenamefont {Rudner}, \citenamefont {Refael},\ and\
  \citenamefont {Lindner}}]{titum2016}%
  \BibitemOpen
  \bibfield  {author} {\bibinfo {author} {\bibfnamefont {Paraj}\ \bibnamefont
  {Titum}}, \bibinfo {author} {\bibfnamefont {Erez}\ \bibnamefont {Berg}},
  \bibinfo {author} {\bibfnamefont {Mark~S.}\ \bibnamefont {Rudner}}, \bibinfo
  {author} {\bibfnamefont {Gil}\ \bibnamefont {Refael}}, \ and\ \bibinfo
  {author} {\bibfnamefont {Netanel~H.}\ \bibnamefont {Lindner}},\ }\bibfield
  {title} {\enquote {\bibinfo {title} {Anomalous floquet-anderson insulator as
  a nonadiabatic quantized charge pump},}\ }\href {\doibase
  10.1103/PhysRevX.6.021013} {\bibfield  {journal} {\bibinfo  {journal} {Phys.
  Rev. X}\ }\textbf {\bibinfo {volume} {6}},\ \bibinfo {pages} {021013}
  (\bibinfo {year} {2016})}\BibitemShut {NoStop}%
\bibitem [{\citenamefont {Kundu}\ \emph {et~al.}(2020)\citenamefont {Kundu},
  \citenamefont {Rudner}, \citenamefont {Berg},\ and\ \citenamefont
  {Lindner}}]{kundu2020}%
  \BibitemOpen
  \bibfield  {author} {\bibinfo {author} {\bibfnamefont {Arijit}\ \bibnamefont
  {Kundu}}, \bibinfo {author} {\bibfnamefont {Mark}\ \bibnamefont {Rudner}},
  \bibinfo {author} {\bibfnamefont {Erez}\ \bibnamefont {Berg}}, \ and\
  \bibinfo {author} {\bibfnamefont {Netanel~H.}\ \bibnamefont {Lindner}},\
  }\bibfield  {title} {\enquote {\bibinfo {title} {Quantized large-bias current
  in the anomalous floquet-anderson insulator},}\ }\href {\doibase
  10.1103/PhysRevB.101.041403} {\bibfield  {journal} {\bibinfo  {journal}
  {Phys. Rev. B}\ }\textbf {\bibinfo {volume} {101}},\ \bibinfo {pages}
  {041403} (\bibinfo {year} {2020})}\BibitemShut {NoStop}%
\bibitem [{Note11()}]{Note11}%
  \BibitemOpen
  \bibinfo {note} {The group $\protect \mathbb {R}$ of real numbers is
  universal cover of the group $\pi _1(SO(2))=S^1$.}\BibitemShut {Stop}%
\bibitem [{\citenamefont {Kennedy}\ and\ \citenamefont
  {Zirnbauer}(2016)}]{kennedy2016}%
  \BibitemOpen
  \bibfield  {author} {\bibinfo {author} {\bibfnamefont {R.}~\bibnamefont
  {Kennedy}}\ and\ \bibinfo {author} {\bibfnamefont {M.~R.}\ \bibnamefont
  {Zirnbauer}},\ }\bibfield  {title} {\enquote {\bibinfo {title} {Bott
  periodicity for $z_2$ symmetric ground states of gapped free-fermion
  systems},}\ }\href@noop {} {\bibfield  {journal} {\bibinfo  {journal}
  {Commun. Math. Phys.}\ }\textbf {\bibinfo {volume} {342}},\ \bibinfo {pages}
  {909} (\bibinfo {year} {2016})}\BibitemShut {NoStop}%
\bibitem [{\citenamefont {Bouhon}\ \emph
  {et~al.}(2020{\natexlab{b}})\citenamefont {Bouhon}, \citenamefont
  {Bzdu\ifmmode~\check{s}\else \v{s}\fi{}ek},\ and\ \citenamefont
  {Slager}}]{bouhon2020b}%
  \BibitemOpen
  \bibfield  {author} {\bibinfo {author} {\bibfnamefont {Adrien}\ \bibnamefont
  {Bouhon}}, \bibinfo {author} {\bibfnamefont {Tomas}\ \bibnamefont
  {Bzdu\ifmmode~\check{s}\else \v{s}\fi{}ek}}, \ and\ \bibinfo {author}
  {\bibfnamefont {Robert-Jan}\ \bibnamefont {Slager}},\ }\bibfield  {title}
  {\enquote {\bibinfo {title} {Geometric approach to fragile topology beyond
  symmetry indicators},}\ }\href {\doibase 10.1103/PhysRevB.102.115135}
  {\bibfield  {journal} {\bibinfo  {journal} {Phys. Rev. B}\ }\textbf {\bibinfo
  {volume} {102}},\ \bibinfo {pages} {115135} (\bibinfo {year}
  {2020}{\natexlab{b}})}\BibitemShut {NoStop}%
\end{thebibliography}%
\end{document}